# Measuring Self-Preferencing on Digital Platforms

Lukas Jürgensmeier* and Bernd Skiera**

February 29, 2024

*click here for most recent version*

**Abstract**

Digital platforms use recommendations to facilitate exchanges between platform actors, such as trade between buyers and sellers. Aiming to protect consumers and guarantee fair competition on platforms, legislators increasingly require that recommendations on market-dominating platforms be free from *self-preferencing*. That is, platforms that also act as sellers (e.g., Amazon) or information providers (e.g., Google) must not prefer their own offers over comparable third-party offers. Yet, successful enforcement of self-preferencing bans—to the potential benefit of consumers and third-party actors—requires defining and measuring self-preferencing across a platform. In the context of recommendations through search results, this research contributes by i) conceptualizing a "recommendation" as an offer's level of search engine visibility across an entire platform (instead of its position in specific search queries, as in previous research); ii) discussing two tests for self-preferencing, and iii) implementing them in two empirical studies across three international Amazon marketplaces. Contrary to consumer expectations and emerging literature, our analysis finds almost no evidence for self-preferencing. A survey reveals that even if Amazon were proven to engage in self-preferencing, most consumers would not change their shopping behavior on the platform—highlighting Amazon's significant market power and suggesting the need for robust protections for sellers and consumers.

**Keywords:** Digital Platforms, Amazon, Competition, Antitrust, Search Engines, Digital Markets Act, American Innovation and Choice Online Act.

*Doctoral Student, Department of Marketing, Faculty of Business and Economics,
Goethe University Frankfurt, Theodor-W.-Adorno-Platz 4, 60323 Frankfurt am Main, Germany,
email: juergensmeier@wiwi.uni-frankfurt.de,

**Full Professor, Department of Marketing, Faculty of Business and Economics,
Goethe University Frankfurt, Theodor-W.-Adorno-Platz 4, 60323 Frankfurt am Main, Germany,
phone: +49 69 798 34649, email: skiera@wiwi.uni-frankfurt.de

*Acknowledgments*

The authors thank Rodrigo Belo, Chiara Farronato, Reinhold Kesler, Lennart Kraft, Ulrich Laitenberger, Jan Landwehr, Filippo Lancieri, Karen Marron, Julien Monardo, Thomas Otter, Devesh Raval, Imke Reimers, Joel Waldfogel, and Amit Zac for helpful comments. Additionally, we thank the participants of the Seventeenth Symposium on Statistical Challenges in Electronic Commerce Research 2021, the Interactive Marketing Research Conference 2021, the Paris Conference on Digital Economics 2022, the European Marketing Academy Conference 2022, the Munich Summer Institute 2022, the ZEW Conference on Information and Communication Technology 2022, the ISMS Marketing Science Conference 2022, the SALTY Quantitative Marketing Conference 2022, the Workshop in Information Systems and Economics 2022, the seminar series at Deakin University, Monash University (Australia), and Goethe University Frankfurt (Germany) for valuable feedback on presentations.

*Financial Disclosure*

The authors gratefully acknowledge financial support from the German Research Foundation (DFG) through grant number SK 66/8-1 and from the INFORMS Society for Marketing Science (ISMS) through the Doctoral Dissertation Award 2023.

# Introduction

A digital platform is an online infrastructure that enables the efficient exchange of goods, services, or information—to which we collectively refer as "offers"—between two or more platform actors (e.g., Constantinides, Henfridsson, and Parker 2018; Goldfarb and Tucker 2019). Numerous types of digital platforms exist, covering a wide range of activities such as retailing, trading, sharing, gaming, and dating.

Efficient exchange requires matching between platform actors such as buyers and sellers. However, the number of offers on platforms is usually large. Therefore, platforms add value by recommending suitable offers to buyers, facilitating discovery, and reducing search costs (Pagani 2013; Ursu 2018). Platform recommendations come in a wide range of formats. For example, a platform provider might label particular sellers as "recommended sellers," designate certain offers as the "platform's choice," or rank offers in its search results.

Crucially, these recommendations strongly influence competition and give platform providers significant power. For example, sellers on platforms like the Amazon Marketplace can struggle to sell products if they do not appear near the top of the search results. Similarly, artists on streaming platforms such as Spotify benefit strongly from being included in curated playlists such as "Best of Hip-Hop" or "Today's Top Hits" (Aguiar and Waldfogel 2021).

In addition to facilitating exchanges among third-party actors, many platform providers serve as platform actors themselves. Amazon, for example, the provider of the Amazon Marketplace platform, also offers products on the platform as a seller. In such a case, the incentives of the roles of platform provider and platform actor do not perfectly align, and *self-preferencing*, a potentially problematic form of preferencing, can occur. Self-preferencing refers to a platform provider's mechanism of preferring its own offers over those of competitors (Ibáñez Colomo 2020). Such behavior implies that, all else being equal, an offer supplied by the platform itself receives better treatment (e.g., stronger recommendations) than a third-party offer.

Currently, it is challenging to accurately identify self-preferencing on digital platforms. In particular, we suggest that a platform's mere tendency to recommend its own products to a greater extent than others is not sufficient evidence for self-preferencing, as such a tendency could occur for "justifiable" reasons. That is, it is possible that the provider's offer deserves to receive more recommendations because it is "better"—e.g., of lower price or of higher quality—than competing offers. Thus, it is inappropriate to test for self-preferencing by simply comparing how prominently the platform provider recommends each offer, without taking into account the various attributes that distinguish offers from one another. As we elaborate in what follows, industry and regulators currently lack a reliable method for addressing this challenge (de Streel et al. 2022).

Despite the difficulty in doing so, developing a robust method of testing for self-preferencing is becoming increasingly critical, in light of recent legislation—including the recently enacted *Digital Markets Act* in the European Union and the proposed *American Innovation and Choice Online Act* in the United States—that seeks to prohibit self-preferencing on dominant platforms, as a means of ensuring fair markets (European Parliament 2022; Cicilline 2021). The need for such legislation is supported by media reports (e.g., Mattioli 2019; Yin and Jeffries 2021) accusing high-profile platforms such as Amazon of engaging in self-preferencing. Perhaps most notably, the US Federal Trade Commission, jointly with 17 US States, sued Amazon in September 2023, alleging the platform illegally maintains monopoly power by "[b]iasing Amazon's search results to preference Amazon's own products over ones that Amazon knows are of better quality" (Federal Trade Commission 2023).

A reliable toolkit for testing for self-preferencing on platforms is a basic prerequisite for regulatory efforts to achieve their stated goal of eliminating self-preferencing. We suggest that, in facilitating this outcome, such a toolkit would significantly benefit all stakeholders that engage with digital platforms. In particular, robust regulations would benefit third-party platform actors by giving them confidence that the platform recommends their offers as deserved. Likewise, they would benefit consumers, by enabling them to trust that platforms indeed recommend the most suitable offers. And if either of these stakeholders suspects that self-preferencing is nevertheless occurring on a particular platform, an established method for evaluating noncompliance would provide them with a means of confirming those suspicions—and the grounds for filing a complaint. In line with this rationale, an agreed-upon approach to measuring self-preferencing would benefit the platform providers themselves, by enabling them to prove their compliance with regulations, and to foster trust with consumers and sellers.

In light of these considerations, this paper aims to provide an approach capable of testing whether competition on a platform, particularly through recommendations, suffers from self-preferencing. Our approach entails assessing whether the extent to which the platform provider recommends its own offers is commensurate with what those offers "deserve," given the offers' attributes. Our work is part of a nascent stream of research on methods for detecting self-preferencing, which we discuss in further detail below, highlighting the distinct features of our approach and analysis.

We apply our approach to data from Amazon Marketplace, the largest digital marketplace in the US and the third largest globally, with an estimated gross merchandise value of 475 billion US dollars in 2020 (Statista 2021). In addition to the US-based Amazon.com, Amazon operates country-specific Amazon Marketplace platforms, e.g., Amazon.de in Germany and Amazon.co.uk in the UK (16 such platforms in total, according to Amazon Seller Central 2022). On these Marketplace platforms, Amazon—the platform provider—serves in a dual role, selling its own products alongside those of third-party sellers.

We focus on one type of recommendation, Amazon's built-in search engine, and examine whether self-preferencing occurs through offers' organic search engine visibility—defined as how prominently an offer ranks in the non-sponsored search results across the entire platform.

We preface our analysis with a consumer survey that seeks to establish the stakes of our inquiry, by exploring the extent to which consumers perceive Amazon's search rankings as affecting their decisions. Further, this survey provides an understanding of consumers' basic expectations of whether self-preferencing occurs on Amazon and how it affects their choices and trust in the platform. Notably, this survey reveals that consumers overwhelmingly believe that Amazon engages in self-preferencing, yet that they are unlikely to diminish their number of future purchases on the platform as a result of such self-preferencing. This finding is indicative of Amazon's market power and highlights the urgency of our approach as a tool in the enforcement of protections for consumers and third-party sellers.

Our main analysis consists of two empirical studies comprising over one million daily product-level time-series observations from three country-specific Amazon Marketplace websites (Germany: Amazon.de; France: Amazon.fr; UK: Amazon.co.uk). We aim to assess whether Amazon engages in self-preferencing by providing more than the "deserved" organic visibility in the search results to offers sold by Amazon as compared with other ("third-party") sellers on the platform.

Our two studies rely on two identification strategies, motivated by two different forms of seller competition on Amazon. In Study A, comprising eight product samples (N = 445,042 observations), we focus on identical products sold by different sellers, including the platform provider, i.e., Amazon. We leverage the fact that Amazon, over time, selects different sellers as the default supplier of a specific offer (by featuring the seller in the product's so-called "buy box" or "featured offer") and examine whether Amazon self-preferences by providing higher search visibility to offers when Amazon, as opposed to a third party, is the default supplier.

In Study B, comprising three different product samples (N = 592,742 observations), we focus on private-label products that individual sellers offer exclusively. We examine whether Amazon self-preferences by providing higher search visibility to its own private-label offers (Amazon Basics) than to private-label substitutes offered by third parties.

In both studies, we find almost no evidence of self-preferencing after accounting for a set of attributes—referred to as "unprotected attributes"—that legitimately influence a recommendation. In Study A, after accounting for unprotected attributes such as prices, sales ranks, product quality, Amazon Prime shipment, and seller rating, we find a statistically significant association between search visibility and the identity of the default supplier (Amazon versus third party) in only one of our eight samples. Specifically, among best-selling products on Amazon.fr (France), Amazon prefers its own products through a 4.9% visibility advantage over third-party suppliers. In the remaining seven samples, including comparable ones from the Amazon Marketplaces in the United Kingdom and Germany, the self-preferencing estimates are not statistically significantly different from zero.

In Study B, after accounting for unprotected attributes that might influence visibility, we find that Amazon Basics products are significantly *less* visible in organic search than comparable close third-party competitors (46.0%, 48.2%, and 49.5% less visible in samples from France, UK, and Germany, respectively). While there is a controversy about whether self-preferencing private-label products would be anti-competitive (e.g., Khan 2019) or not (because it is similar to standard private-label offline retailing (e.g., Dubé 2024)), the results of our empirical studies preempt this debate by not finding evidence for self-preferencing of Amazon's private-label products through organic search results.

We note that, in our empirical setting (Amazon Marketplace), we analyze non-personalized search results gathered through anonymous browsing sessions. On the one hand, this choice makes it possible for our data provider to conduct the approximately one million daily search requests to compute a visibility market share (using logged-in user accounts or a single device to conduct the searches would lead to denied access). On the other hand, using data from anonymous searches adds the benefit that we do not need to consider consumer characteristics (such as past purchasing history that indicates a consumer's preferences) to obtain a valid self-preferencing estimate. While our results cannot speak to whether self-preferencing occurs in personalized search results, our approach readily extends to such settings: researchers can measure potentially personalized recommendations at a smaller scale, for example, via a browser plugin that tracks individual consumers' search results (Farronato, Fradkin, and MacKay 2023) and additionally account for consumer-specific data obtained through surveys or consumers' personal data requests with the platform.

**Previous Literature on Detecting Self-Preferencing on Digital Platforms**

Extant literature acknowledges that testing for self-preferencing is challenging, given that platforms closely guard their algorithms, and publicly available data might be insufficient to capture the considerations informing recommendation decisions. Although recent literature has started considering this challenge, there is currently no firmly established approach to test for self-preferencing on platforms (de Streel et al. 2022).

Several recent studies have empirically identified cases in which digital platforms seemed to give preferential treatment to offers possessing specific attributes. For example, Hunold, Kesler, and Laitenberger (2020) showed that online travel agencies punish actors offering lower prices on other channels through worse search rankings.

Cure et al. (2022) found that meta-search platforms prioritize offers from sales channels with which they are affiliated (owned by the same or linked entities). In contrast to the current research, these works solved a specific problem but did not seek to develop a more general approach for identifying preferencing (self-preferencing or otherwise).

More directly related to our research, the media report of Jeffries and Yin (2021) suggests that Amazon's private label brands often receive the top spot in the search results, albeit without considering the offer's key attributes, such as prices. Considering such attributes, Farronato, Fradkin, and MacKay (2023) used individual consumers' search results to identify whether Amazon tends to provide higher rankings to its own brands (e.g., Amazon Basics). The authors found that, for some keywords, Amazon seems to rank its own brands higher than those of competitors, even when controlling for observable attributes such as price, rating, and delivery options. That research considers search results ranks of small sets of keywords; our work proposes a more holistic metric of the extent to which a platform recommends a particular offer: *search engine visibility*, capturing the visibility market share of individual products in the entire marketplace. Hence, our results can better gauge whether a platform engages in self-preferencing across the entire platform.

Aguiar, Waldfogel, and Waldfogel (2021), focusing on the music streaming industry, assessed platform ranking bias on Spotify by measuring offers' (here: songs') levels of ex-post success, conditional on the recommendation (i.e., the ranks) the platform assigns them in a playlist. The assumption underlying this approach is that platform recommendation should match the extent to which consumers actually "like" (i.e., derive utility from) the offers, and that any discrepancy between the recommendation and consumers' "liking" of an offer is an indicator of bias.

In a recent working paper, Reimers and Waldfogel (2023) suggested that the works of Farronato, Fradkin, and MacKay (2023) and Aguiar, Waldfogel, and Waldfogel (2021)—together with an earlier version of this article—exemplify two main tests for identifying (self-)preferencing in digital platforms. The first test, represented by an earlier version of this article and by that of Farronato, Fradkin, and MacKay (2023), is referred to as "Conditioning-on-Observables" (COO), which entails "[regressing] ranks on controls and a platform indicator, and interpreting the coefficient on the platform indicator as bias" (Reimers and Waldfogel 2023, p. 11). Similarly, Raval (2022) used offer attributes to explore whether Amazon preferences itself when selecting the default seller (the "winner of the buy box") for a particular product. The other test described by Reimers and Waldfogel (2023), exemplified by Aguiar, Waldfogel, and Waldfogel (2021), is referred to as "outcome-based" (OB), as it evaluates preferencing on the basis of products' ex-post success. (The principles underlying the two tests are discussed further in the "Description of Conceptual Framework" section below.)

Reimers and Waldfogel (2023) compared the COO and OB tests, applying each test to e-book's rankings on Amazon's Kindle Daily Deal page to identify whether Amazon assigns higher rankings to e-books published with its own imprint (vs. third-party publishers). In their comparison, the authors highlighted the comparatively low and easy-to-obtain data requirements as a key advantage of the COO test and the potential of omitted variable bias as a disadvantage. We contribute to the debate by implementing the COO and OB tests to visibility across the entire Amazon marketplace, intensively discussing the omitted variable problem (of both tests), and suggesting solutions to rule it out with a combination of regulatory disclosure requirements, a consumer survey, and a robustness check through the OB test.

Together, the studies cited above—many of which are still works in progress—highlight the nascent nature of research on self-preferencing on digital platforms, and the urgent need for robust methodologies for identifying self-preferencing behavior when it occurs.

## Relevance of Self-Preferencing for Platform Stakeholders

To motivate our research, we first provide a basic "sketch" of the expected effects of platform self-preferencing on the stakeholders that interact with the platform. For more formal (theoretical) examinations of what happens when a platform preferences some offers over others, see, for example, Barach, Golden, and Horton (2020) and Long and Amaldoss (2024).

Figure 1 illustrates a simplified consumer choice process on a digital platform and shows how the platform provider's recommendation affects the involved stakeholders. Consumers use the platform to search for offers. A platform enables actors to supply their offers. Those platform actors can be first- and third-party actors. A first-party actor is affiliated with the platform provider, while third-party actors are independent entities. For example, the first-party actor Amazon sells products on its own platform that compete with products supplied by third parties.

**Figure 1:** Stakeholders on a Digital Platform and Their Role in the Consumer Choice Process

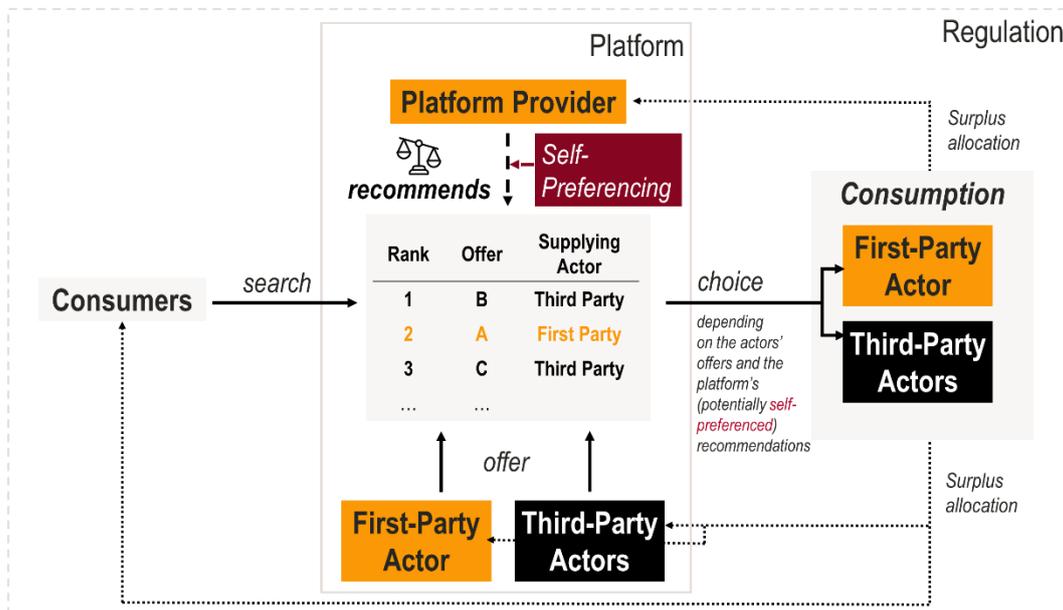

When a consumer browses the platform, the platform provider guides this process by providing recommendations for specific offers—each supplied by a particular actor. A key means of recommending offers is through organic search results, generated in response to search terms (also known as keywords) entered by the consumer. The central box in **Figure 1** illustrates such search results. The search results inherently present the offers in a particular order, corresponding to the extent to which the platform provider "recommends" each offer to consumers.

Consumers now choose among the recommended offers, and we can think of the choice as a function of the offer characteristics, the platform's recommendation, and—potentially—self-preferencing. This choice now generates a surplus allocated towards the stakeholders involved: the platform provider, first- and third-party actors, and consumers.

Because self-preferencing affects consumers' choices directly, it also affects surplus allocation. Intuitively, the platform provider could skim the surplus by steering consumers to first-party actors' offers. This distortion in marketplace outcomes could decrease consumers' and third-party actors' surplus by steering consumer choice towards lower-utility first-party offers.

Lawmakers and regulators often label such conduct as "unfair competition" (e.g., Federal Trade Commission 2023) or "unfair practices" (e.g., European Parliament 2022), particularly if platforms are market-dominating. Therefore, regulators would like to know whether self-preferencing abuses the platform provider's dominant position by steering consumers to the platform's own offers.

Because of this regulatory scrutiny, platform providers care deeply about whether their recommendations are self-preferencing. As discussed above, lawmakers are enacting laws prohibiting self-preferencing recommendations with significant fines for violations—the recently enacted Digital Markets Act in the European Union, for example, stipulates fines of up to 10% of a platform provider's *annual worldwide revenue* if a platform provider self-preferences (European Parliament 2022, article 30, paragraph 1), making compliance a commercial necessity. We suggest that means of evaluating self-preferencing, such as the approach presented herein, would enable all stakeholders—consumers, third-party actors, regulators, and the platform itself—to share a common understanding of what compliance entails, to the joint benefit of all.

**Description of Conceptual Framework**

Building on Reimers and Waldfogel's (2023) categorization of tests for identifying self-preferencing, Figure 2 illustrates our conceptual framework. Specifically, it shows how the two tests—Conditioning-on-Observables (COO) and the Outcome-based (OB) test—relate to each other. In effect, both approaches aim to estimate the impact of a *protected attribute*, that is, an attribute that should not be used to determine an offer's recommendation, on a dependent variable of interest. In our setting, which explores self-preferencing, the protected attribute indicates whether or not the platform supplies a focal offer (denoted *isPlatform*).

**Figure 2:** Conceptual Framework with Two Approaches to Test for Self-Preferencing Through Recommendations

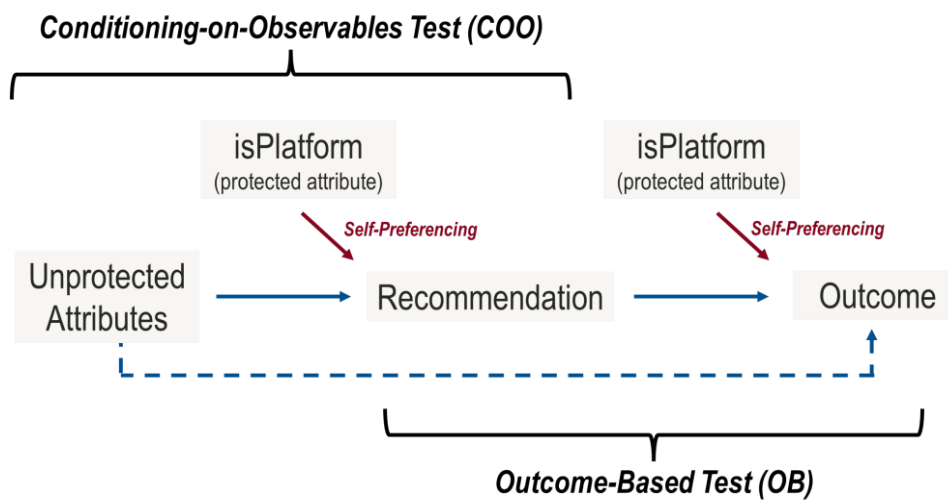

In the COO test, the primary focus of this manuscript, the dependent variable of interest is the *recommendation* itself. Here, the recommendation is modeled as a function of unprotected attributes (i.e., attributes deemed relevant for a recommendation, such as price or rating) together with the protected attribute. If this protected attribute influences the recommendation, we can determine that the recommendation suffers from self-preferencing.

In the OB test, in turn, the dependent variable of interest is the *outcome* of the consumer journey (e.g., the purchase decision or sales quantity). Here, the outcome is modeled as a function of the recommendation and the protected attribute. This test's key idea is that an offer should receive the same outcome conditional on the recommendation, irrespective of whether the offer belongs to the platform (i.e., the value of *isPlatform*). If the platform offer received the same outcome with a better recommendation than the third-party offer, then the recommendation would suffer from self-preferencing.

The two tests have different advantages and disadvantages. A key advantage of the OB test over the COO test is that it does not require analysts to observe and control for all unprotected attributes that influence the outcome through the recommendation, because the recommendation itself already accounts for those unprotected attributes. At the same time, the OB test necessitates knowledge of the outcome resulting from a recommendation (e.g., a purchase decision)—as well as a means of verifying that the outcome is indeed a causal consequence of the recommendation (e.g., the consumer purchased *because* of the recommendation). In other words, it is necessary to control for other attributes that influence the outcome through paths other than the recommendation while being correlated with the protected attribute (highlighted by the dashed line in Figure 2). It is difficult to satisfy these requirements without having access to platform-internal data (and even then, it might be challenging).

Table 1 defines the constructs of the conceptual framework to measure self-preferencing, outlines common measurements, and highlights extant literature that uses those measurements to identify self-preferencing on e-commerce platforms. Additionally, we show how our study builds on and differs from existing literature on measuring self-preferencing.

**Table 1:** Definition of Constructs and their Potential Measurement

| Construct<br>*Definition* | Potential Measurement | Extant Literature | This Study |
|---|---|---|---|
| **Recommendation**<br>*The prominence of an offer on a platform, determined by the platform provider.* | **Top result**: frequency of an offer appearing as the first search result. | "The Markup" newspaper article (Jeffries and Yin 2021) | — |
| | **Rank**: an offer's rank in individual search results. | FFM (Farronato, Fradkin, and MacKay 2023); HKL (Hunold, Kesler, and Laitenberger 2020); Cure et al. (2022), RW (Reimers and Waldfogel 2023). | Input to visibility. |
| | **Featured Offer:** mechanism recommending an offer's default supplier (e.g., the Amazon Buy Box) | Raval (2022). | Input to identification of self-preferencing |
| | **Visibility**: an offer's search engine visibility across the platform (i.e., visibility "market share") | — | ✓ |
| **Outcome**<br>*Consumers' action after the recommendation.* | **Consumer Choice:** binary choice from a choice set, e.g., clicks or purchases. | RW (2023). | — |
| | **Sales (Ranks):** number of product sales (or sales rank within a category). | FFM (2023), HKL (2020), Raval (2022), Cure et al. (2022), RW (2023). | ✓ |
| **Unprotected Attribute**<br>*Offer attributes that can inform the platform provider's recommendation.* | *How to identify attributes?* | | |
| | **Observables**<br>Observable offer attributes (+ fixed effects), excluding protected attributes. | All other previously mentioned studies. | ✓ |
| | **Regulatory Disclosure**<br>Platform-disclosed attributes influencing recommendation based on regulation. | — | ✓ |
| | **Consumer Survey**<br>Survey to identify which attributes provide value for consumer choice. | — | ✓ |
| **Protected Attribute**<br>*Offer attributes that should **not** inform the platform provider's recommendation.* | **Regulation**<br>Regulation that identifies protected attributes (e.g., *isPlatform* due to prohibition of self-preferencing). | All other previously mentioned studies. | ✓ |
| | **Consumer Survey**<br>Identifying attributes that do not provide value to consumer choice. | — | ✓ |

**Consumer Perspective: Self-Preferencing on Amazon**

Given that our empirical investigation focuses on Amazon, we sought to further establish the stakes of our inquiry by gaining additional direct insights into consumers' perceptions of the extent to which Amazon engages in self-preferencing, and how such self-preferencing might affect consumers' behavior. We also used the survey to understand which offer attributes guide consumers' choices on Amazon. To this end, we surveyed N = 300 frequent online shoppers from the United Kingdom, recruited through Prolific in January 2024. All participants had shopped on

Amazon.co.uk at least once in the previous year. Participants' ages ranged from 19 to 77 years (mean = 39.9 years, SD = 12.9 years). The sample was gender-balanced (50% female), and the median consumer placed an order on Amazon more than once per month—ensuring a certain degree of familiarity with the platform.

In response to our survey, the overwhelming majority of consumers stated that the search results ranking on Amazon affects their product choice. While the median consumer indicated that the rankings influence their choice "much" (4 on a 5-point Likert scale), only 5.6% of consumers claimed that the rankings do not affect their choice at all. This result corroborates findings from the literature showing that the search result ranking matters for consumer choice (e.g., Agarwal, Hosanagar, and Smith 2011; Drèze and Zufryden 2004; Ghose, Ipeirotis, and Li 2014). Because self-preferencing changes the ranking compared to no self-preferencing, and the choice depends on the ranking, we can infer that self-preferencing is likely to affect consumer choice.

We also found that almost all surveyed consumers know that third parties supply products on Amazon (96%). However, they tended to overestimate the proportion of products that Amazon supplies; the median consumer estimated that Amazon supplies 61% of all products on the platform, though estimates varied widely (SD = 21 percentage points). In fact, our analysis of data from Keepa indicates that Amazon only supplies one-fourth of all currently available products on the British marketplace in January 2024.

We also asked participants to indicate which attributes influence their purchase decisions. For this question, we presented consumers with a list of all attributes disclosed by Amazon to influence the recommendation and additional attributes that could play a role. Not surprisingly, price influences choices: 86% of consumers indicate that it influences their choices. With 80% agreeing, product reviews are the second-most frequently mentioned attribute, followed by Prime shipment (58%), seller reviews (54%), product's brand (28%), how well the product sold in the past (26%), the product's in-stock percentage (11%). Only 9% of consumers indicate that the seller's identity influences their choices (beyond the associated quality of the seller, which is assessed separately through the seller reviews). Similarly, few consumers (9%) indicate whether a product is an Amazon private label product influences their choices. These low values suggest that—beyond a regulatory requirement—the mere seller identity (apart from the associated product quality)—is not a relevant choice criterion for most consumers.

Regarding self-preferencing, the vast majority of consumers reported suspecting that self-preferencing occurs on Amazon—78% of respondents think that Amazon prefers products they supply in the search results over the same product when a third party supplies it. Hence, consumers' expectations—of how rankings affect their choices and if self-preferencing occurred—align with the sentiment in the press and the premise underlying a ban on self-preferencing (that self-preferencing could occur and that it changes consumers' choices).

Our survey results further indicated that self-preferencing is likely to affect consumers' trust in the platform. Specifically, we asked how consumers would change their trust in the platform if an investigation found self-preferencing, and 37% stated they would decrease their trust in the platform (52% would not change, and the remaining participants did not know or indicated increased trust). Conversely, when asked how they would respond if the investigation found no self-preferencing, 41% stated that they would increase their trust in the platform (58% would not change, and 1% would decrease their trust).

Notably, despite the substantial effects of self-preferencing on trust, only a minority of consumers indicated that evidence of self-preferencing would change their purchasing behavior. Specifically, when asked how they would change their purchase behavior if an investigation found evidence for self-preferencing, only 2% of consumers stated that they would decrease their purchases on Amazon very much, and only 13% indicated they would decrease their purchases slightly—whereas 74%, the vast majority of consumers, indicated they would not change how much they purchase on Amazon. Conversely, if Amazon were found not to self-preference, 13% would increase their number of purchases slightly (3% very much), and the overwhelming majority would continue to purchase as much as before (82%).

Together, these survey results suggest that while self-preferencing is expected to influence consumers' trust in Amazon, it might not affect consumers' actual purchase behavior in most cases. This finding might be attributable to Amazon's considerable market power—indicating that the platform could use self-preferencing to steer consumers to first-party offers while suffering minimal negative consequences in terms of how much consumers purchase on the platform.

### Context and Approach of the Empirical Studies

In what follows, we lay the foundations for our identification strategy by describing the specific mechanisms of seller competition in our empirical context, the Amazon Marketplace. We subsequently describe a five-step approach for implementing what (Reimers and Waldfogel 2023) refer to as the COO test for self-preferencing on a digital platform, and show how we apply each step in our empirical setting.

#### Two Forms of Seller Competition on the Amazon Marketplace

This research examines two main forms of seller competition on the Amazon Marketplace: competition *within* products—in which multiple sellers offering the same product compete to be the default seller (addressed in Study A); and competition *between* products—in which sellers compete over different products (addressed in Study B). Figure 3 illustrates these two forms of competition.

**Figure 3:** Two Forms of Offer Competition Motivating Two Empirical Studies

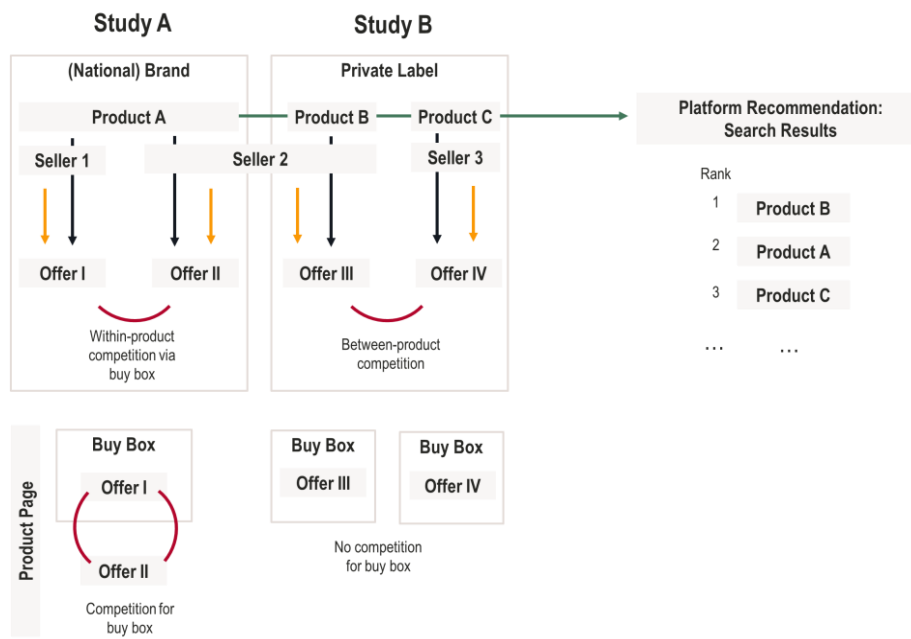

*Within-product competition through the buy box (Study A)*

Often, many sellers compete to sell a specific product. In such cases, Amazon does not list each offer individually (e.g., in the search results) but unifies those offers into one listing. Sellers now have to compete for sales of that product with other sellers via the buy box. The buy box is a visually separated box on the product page's top-right corner, highlighting one seller as the default seller, also referred to as the "featured offer." That seller might be Amazon or a third-party seller, and it can vary over time. Most customers proceed to purchase the product from the recommended buy box seller. Indeed, though Amazon does not communicate the exact share, industry experts estimate that the buy box seller receives between 80% (Lanxner 2019) and close to 90% (Zeibak 2020) of product sales. Accordingly, the buy box creates a situation in which, at any given time, each product is effectively supplied by only one seller: the seller who holds the buy box.[1]

Amazon uses a proprietary algorithm to determine the buy box seller. The company's third-party seller documentation names four attributes influencing the probability of winning the buy box (which do not necessarily correspond to the attributes influencing the search results ranking): competitive pricing, fast and free shipping, high-quality customer service, and a high in-stock percentage (Amazon

---

[1] Amazon distinguishes between buy boxes for new and used products. In our research, we only analyze the seller of the buy box for new products because the "used" buy box occurs relatively infrequently (a spot check of Keepa's Amazon.de database in February 2024 shows that only 2.9% of all currently available products have a "used" buy box). Still, one could compare varying conditions of used products to new products by additionally taking the products' used condition into account.

Seller Central 2022). Similarly, Reiner, Rutz, and Skiera (2023) find that the shipping method (e.g., whether Amazon fulfills product orders) and the lowest price influence the buy box decision. They also find that the buy box seller changes, on average, every 82 hours.

In Study A, we leverage this variation in the buy box seller over time to identify whether a focal product is more visible in the search results when Amazon holds the buy box than when a third party holds the buy box.

*Between-product competition through private-label products (Study B)*

Sellers can avoid competition within products by offering private-label products for which they are the exclusive seller. One of Amazon's most prominent examples of a private-label brand is "Amazon Basics." Under this label, Amazon exclusively sells a range of private-label products. For Amazon Basics products, Amazon does not have to compete for the buy box but instead competes with other (usually third-party) products. Hence, competition for private-label products is *between* products instead of *within* products via the buy box. Accordingly, we will compare Amazon Basics products' visibility with that of comparable third-party private-label products in Study B.

***Five-Step Approach to Test for Self-Preferencing on the Amazon Marketplace***

In what follows, we describe a five-step approach that implements the COO test for self-preferencing through recommendations. Table 2 outlines the five steps of our approach in our empirical studies on the Amazon Marketplace.

**Table 2:** Five-Step Approach to Test for Self-Preferencing and its Implementation in Our Studies on the Amazon Marketplace

| | **Step of Approach** | **Implementation in our Study** |
|---|---|---|
| 1 | Measure the degree of recommendation. | Measure organic search visibility; see equations (1) to (6). |
| 2 | Identify and measure attributes that could influence the recommendation. | Identify attributes that could influence the recommendation by i) starting with observable attributes, then ii) cross-checking those with Amazon's disclosure of attributes influencing the search results in response to EU regulation 2019/1150 (European Parliament 2019), and iii) verifying those matched attributes through a consumer survey of which attributes inform consumers' choices. |
| 3 | Define self-preferencing in the specific context. | *Definition of self-preferencing:* Amazon's search engine prefers Amazon-supplied products over third-party-supplied products, conditional on unprotected attributes. |
| 4 | Categorize attributes into protected and unprotected attributes. | Unprotected attributes are product attributes that inform consumer choice and are disclosed by the platform, excluding protected attributes. |
| | | *isPlatform* is a protected attribute that should not influence the recommendation following laws prohibiting self-preferencing, such as the Digital Markets Act (European Parliament 2022). |
| 5 | Estimate the recommendation as a function of protected and unprotected attributes. Test $H_0$: $\beta_{protected} = 0$. | Two-way fixed effects generalized linear model with log-link, see Equation (7). Assess confidence intervals for the likelihood of type-1 error and robustness check for influence of type-2 error by excluding or accounting for days on which measurement error could occur most likely. |

*Step 1: Measure the degree of recommendations through organic search engine visibility.*

Step 1 of the five-step approach measures the degree of recommendation of individual offers. Our analysis focuses on recommendations in the form of an offer's rankings in a platform's organic search engine results. Rather than using the raw search results rankings as previous literature does, we build on these rankings to propose a more holistic construct of *visibility*—which reflects an offer's market share of search engine visibility on the entire platform.

Our definition of visibility follows Sistrix (2021), a search engine optimization firm that popularized the metric and supplies visibility data for this study. The core metric is the visibility of offer $i$ in period $t$, $V_{it}$. It is the sum of the "raw" visibility $V_{kit}$ of offer $i$ for keyword $k$ in period $t$ across all keywords $k \in K$, where $K$ is a given set of keywords used on the platform:

$$V_{it} = \sum_{k \in K} V_{kit} . \tag{1}$$

To compute the value of $V_{kit}$, we multiply the number of consumers' search queries $N_{kt}$ for keyword $k$ in period $t$ by a function that depends on the ranking of offer $i$ in the search results obtained from a search query with keyword $k$ in period $t$:

$$V_{kit} = N_{kt} \times f(\text{rank}_{kit}) \qquad (2)$$

The function $f$ can be defined in various ways; one straightforward way of defining it is as an inverse relationship between the search rank and the expected click probability (ECP). In other words, the greater a listing's search rank (meaning, the further down it appears in the search results), the lower the probability of clicking on the listing. The ECP (differently from, e.g., an offer-specific observed click-through rate), estimates how likely consumers are to click on an offer on a given rank—across similar keywords and irrespective of which offer occupies this rank.

We prefer the ECP over the raw ranks because, on the one hand, the ECP incorporates the empirical finding that consumers' probability of clicking on a search result concentrates disproportionally among the top search results and decreases rapidly with increasing rank (e.g., Agarwal, Hosanagar, and Smith 2011; Feng, Bhargava, and Pennock 2007; Ghose and Yang 2009; Skiera and Nabout 2013). On the other hand, it leaves enough flexibility to account for different functional forms of the expected "rank–click" relationship across different offer categories or search intents. Still, and importantly for our application, the ECP is only rank- and keyword-specific, but not offer-specific. Hence, the ECP does not measure the offer's success but instead characterizes how "good" a given rank is for a given keyword, irrespective of which specific offer occupies this rank.

For our studies, the data provider Sistrix records the search results in anonymized browsing sessions, i.e., without any user information or cookie retention. Sistrix estimates $f$ based on third-party data for the number of impressions and clicks for keyword-specific ranks in the search results.[2] Accordingly, our variable accounts for the highly non-linear ECP for each rank. Weighting the ECP by the number of search queries for a keyword adds the benefit of making the search results corresponding to different keywords comparable, thereby enabling the eventual computation of a visibility "market share" on the entire platform.

Because our data are derived from non-personalized searches, using our metric enables us to draw conclusions on the search results absent personalization and, thus, absent any expectations the platform might have about a specific user's behavior. Nevertheless, our approach would also work with personalized ranking data, for example, if the analyst can obtain a sufficiently large number of personalized search

---

[2]Trade secrets prevent the firm from disclosing the exact procedure to estimate $f$. However, Beus (2015) and Beus (2020) outline Sistrix's general procedure for estimating the ECP on Google. Sistrix assured us that they use a similar method for estimating the ECP on Amazon.

results (as a dependent variable) and user characteristics (as additional independent variables).

The visibility metric $V_{kit}$, defined in equation (2), depends on seasonal search traffic fluctuations. That is, since $V_{kit}$ depends multiplicatively on $N_{kt}$, its value is higher when more users search for a particular keyword, even if it just represents a seasonal increase. Hence, comparing an offer's search rankings at different points in time creates a risk of conflating (seasonal) changes in the popularity of an offer with changes in the platform's rankings of that offer. As we are interested only in the latter changes, we account for seasonality by weighting $f$ by the seasonality-free popularity of a given search query.

To this end, we replace $N_{kt}$ in equation (2)—i.e., the number of searches in period $t$—with the number of average keyword searches over the entire length of the past seasonal cycle:

$$\bar{N}_{kt} = \frac{1}{|M|} \sum_{m \in M} N_{k,\ t-m+1}, \qquad (3)$$

where $|M|$ is the length of the cycle. In our case of a yearly cycle and daily observations, the cycle length is $|M| = 365$ days.

Thus, replacing $N_{kt}$ from equation (2) with $\bar{N}_{kt}$ defined in equation (3), we compute the seasonality-free search query visibility index for offer $i$ given the keyword search $k$ in period $t$ as follows:

$$\text{VI}_{kit} = \bar{N}_{kt} \times f(rank_{kit}). \qquad (4)$$

Next, similarly to equation (1), we aggregate this keyword-level visibility index over the entire set of keywords $K$ on the platform:

$$\text{VI}_{it} = \sum_{k \in K} \text{VI}_{kit}. \qquad (5)$$

As the last computation, we transform the offer-specific visibility index $VI_{it}$ into a *relative* visibility index. This metric sets the individual offer's visibility in relation to the sum of all offers' visibility in the same period $t$

$$\widetilde{VI}_{it} = \frac{\text{VI}_{it}}{\sum_{i \in I} \text{VI}_{it}}, \qquad (6)$$

where $I$ is the set of all offers on a platform. Thus, $\widetilde{VI}_{it}$—the relative visibility index of an offer $i$ in period $t$—is a metric of an offer's search engine visibility that (i) is free of seasonal search traffic influences and (ii) is measured relative to the total visibility of all platform offers. Thus, the relative visibility index, $\widetilde{VI}_{it}$, is a metric representing each offer's share of total visibility in each period. In the following, we use the term *visibility* for the relative visibility index, defined in equation (6).

Sistrix computes each product's visibility as described in equations (1) to (6). Because the Amazon Marketplace features millions of offers, an individual offer's visibility share is very small. Hence, Sistrix (2020) multiplies those values by one million for ease of interpretation. The metric builds on data from an automated daily web scraper that records the results of one million representative keyword searches on the Amazon search engine. We use Sistrix's daily visibility metric for each product as the dependent variable in our analyses.

*Step 2: Measure attributes influencing visibility*

Table 3 displays the attributes that influence organic search engine visibility on the Amazon marketplace, according to Amazon's disclosure due to EU regulation 2019/1150 (European Parliament 2019). Consumers can access this information on Amazon marketplaces falling under the EU's jurisdiction by clicking "learn more about the results" at the top of the search results. We map those disclosed attributes to the observable variables or, in case those are unobserved, explain how we account for them in the model. For the unbiased estimation self-preferencing in Step 5, we assume that no additional attributes that correlate with the protected attribute influence the recommendation. We additionally verified this assumption by confirming in our survey which attributes do and do not inform consumers' choices; notably, this survey also enabled us to ensure that no important attribute was omitted from our analysis. Further, Table 3 compares our choices to those used in the three most closely related studies using COO-based approaches to test for self-preferencing.

We collected data from three country-specific Amazon Marketplace websites, corresponding to the three largest economies by Gross Domestic Product in Europe: the German (Amazon.de), British (.co.uk), and French (.fr) websites. Specifically, we collected daily product-level observations, corresponding to 5,503 different products, during the period ranging from May 27, 2020, to January 23, 2021 (daily time series panel data with $T = 242$ days; $N = 1,037,784$ observations). This period includes the coronavirus pandemic, which greatly affected demand for e-commerce. However, our analysis focuses on relative differences between offers supplied by Amazon versus third parties, and we assume that the broader trends in e-commerce demand did not affect these relative differences. Additionally, we made sure to only include products that were available for the majority of the observation period.

We obtained data on all attributes influencing visibility (for use as independent variables in Step 5) from the commercial Amazon tracker Keepa.com. Keepa is a data provider with a unique record of almost all available Amazon products in several important international markets. The firm updates its database multiple times daily and stores the historical values for each product.

The data we collect from Keepa for each product and each day include the following: the price charged by each seller that offers the product; the product's buy box seller on a focal day; the product's sales rank within its broadest category, the number of reviews for the product, the product's average star rating, the rating for each seller offering the product, and whether the buy box seller offers Amazon Prime shipping for the product.

**Table 3:** Amazon's Self-disclosed Attributes Influencing Search Rankings, Operationalization in this Study, and Comparison with Other Studies.

| Attribute influencing Amazon's default "featured" search ranking, as disclosed by Amazon (2022) | Operationalization in this study | Share of consumers stating that this attribute influences their choice | Inclusion of attributes in other self-preferencing analyses of search results on the Amazon Marketplace | | |
|---|---|---|---|---|---|
| | | | Jeffries and Yin (2021) *Aim:* Predicting the top search result is an Amazon private label product. | Farronato, Fradkin, and MacKay (2023) *Aim:* Identify self-preferencing of Amazon private labels vs. non-Amazon products. | Reimers and Waldfogel (2023) *Aim:* identify self-preferencing of Amazon-related e-books on the Kindle daily deal page. |
| "Customer actions, e.g., how often an item was bought" | Sales rank | 26% | ✗ | ✗ | ✓ |
| "Information about the item, e.g., title, price, and description" | Price, | 86% | ✗ | ✓ | ✓ |
| | rating, | 80% | ✗ (✓ in auxiliary analysis) | ✓ | ✓ |
| | product-fixed effects | | ✗ | ✗ (impossible because of study design) | ✗ (impossible because of study design) |
| "Delivery speed" | isPrime (indicating whether the offer is available for fast Prime shipment) | 58% | ✓ (indicator if shipped by Amazon) | ✓ (delivery time) | (not necessary because digital good) |
| "Availability" | Sample only includes available products | 11% | Unclear | ✓ (control for availability) | (not necessary because digital good) |
| "Costs, e.g., shipping costs" | Price, including shipping fees | 86% (see above) | ✗ | ✓ | ✓ |
| "Whether we think the item will be of interest, e.g., new items." | • Product- (Study A) or comparison-group fixed effects (Study B), | | ✗ | ✗ (similarity between keyword and product title) | ✗ |
| | • sample only includes established products | | unclear | ✗ | ✗ (but different context) |

Keepa checks time-varying product data multiple times per day but only records a value once a variable changes. We use the last observation of that day to transform such irregularly-spaced time-series data into daily observations. In contrast, Sistrix only provides a single value per day. Hence, it could be possible that Sistrix records a product's visibility on day $t$ before a product's price, sales rank, or buy box seller changed at the end of day $t$. To avoid such a situation, we lag all independent variables from Keepa by one day, which amounts to a time difference of a few hours. This difference also ensures that the search ranking has already been updated and reflects potential product attribute changes. Our robustness checks in Web Appendix B ensure that the small measurement error introduced by non-identical measurement times during the day is unlikely to affect our conclusions.

*Step 3: Define self-preferencing*

The Digital Markets Act prohibits self-preferencing for dominant platforms, so-called gatekeepers: "The gatekeeper shall not treat more favourably, in ranking and related indexing and crawling, services and products offered by the gatekeeper itself than similar services or products of a third party." (European Parliament 2022, article 6, paragraph 5). In line with this definition, we define self-preferencing as a situation in which Amazon's search engine provides higher rankings to Amazon-supplied products than to third-party-supplied products, conditional on unprotected attributes.

*Step 4: Categorize attributes into protected and unprotected attributes.*

Based on this reasoning, we classify the seller's name (here, the indicator if Amazon or a third party supplies the offer) as a protected attribute. We consider the publicly disclosed attributes influencing visibility (see Step 2 and Table 3) unprotected attributes since laws do not explicitly prohibit them from influencing the recommendation. Note that we include the seller rating as a proxy for the seller's quality. Hence, the following model only attributes the effect of the seller name, not seller quality, on the recommendation.

*Step 5: Estimate Recommendation = f(Unprotected, Protected).*

Finally, Step 5 requires us to model the recommendation, measured by visibility, as a function of the unprotected and protected attributes described above. We use the following two-way fixed effects Generalized Linear Model (GLM) with log-link (also referred to as "Poisson regression") to model the expected search engine visibility conditional on and as a function of the following protected and unprotected attributes:

$$E\left[\widetilde{VI}_{it} \middle| \cdot \right] = \exp\left(\delta \text{ isAmazon}_{it} + \beta_1 \ln(\text{salesRank}_{it}) + \beta_2 \ln(\text{price}_{it}) \right. \quad (7)$$
$$+ \beta_3 \ln(\text{countReviews}_{it}) + \beta_4 \text{ ratingProduct}_{it}$$
$$\left. + \beta_5 \text{ ratingSeller}_{it} + \beta_6 \text{ isPrime}_{it} + \alpha_i + \gamma_t \right),$$

where the protected (binary) attribute *isAmazon* = 1, if Amazon holds product $i$'s buy box on day $t$ (Study A) or if the product $i$ is an Amazon Basics product (Study B). *salesRank* is a proxy for sales and refers to the product's sales rank in its broadest category, *price* is the current buy box seller's price (including shipping fees), *ratingProduct* is the product's rating in stars, ranging from 1 to 5, and *isPrime* is an indicator that equals 1 if the product is available for Amazon Prime shipment. Amazon-supplied products are usually available through Amazon Prime, which third-party sellers can also achieve through "Fulfillment by Amazon" (Lai et al. 2022).

Importantly, we include *ratingSeller*, which captures the effect of seller quality on the recommendation. Hence, $\hat{\delta}$ estimates the mere effect of being Amazon on the recommendation. In Study A, $\alpha_i$ refers to the product-fixed effect, while Study B accounts for comparison-group-fixed effects, including the Amazon Basics and up to five similar competing private-label products. $\gamma_t$ are date-fixed effects.

We note that the correct identification of $\hat{\delta}$ requires that we account for all unprotected attributes—observed and unobserved—that influence an offer's recommendation and could correlate with the recommendation. Failing to do so could lead to omitted variable bias. Hence, as elaborated in Step 2, we rely on EU regulation 2019/1150 (European Parliament 2019), which requires platforms to make attributes influencing the recommendation transparent. Additionally, we validate these attributes by asking consumers which attributes influence their product choice. If we account for all attributes influencing the recommendation, as shown in Table 3, and a platform provider does not use unpublished and unaccounted attributes that correlate with the protected attribute, a statistically significant positive estimate of the protected attribute's coefficient would indicate self-preferencing.

Next, we implement the above considerations in the specific setting and introduce our two empirical studies, each corresponding to one of the two main forms of competition on the Amazon Marketplace.

## Study A: Testing for Self-Preferencing Through Search Results Among Products Sold by Multiple Suppliers

Study A focuses on products sold on Amazon Marketplace by multiple suppliers over time. In this study, we analyze product listings for which Amazon and third-party sellers hold the buy box at different times. We use this seller variation *within* products to test whether Amazon's search algorithm prefers Amazon's products by assigning these products higher-than-justified visibility compared to third-party sellers; in other words, whether Amazon self-preferences in the search results. Specifically, we examine whether—given that the product is identical and that we control for other unprotected attributes that might influence search visibility—the product is more visible in the search results on days when Amazon, rather than a third party, holds the buy box.

*Sample Selection and Descriptive Statistics*

Study A uses eight samples comprising daily observations for different samples of products obtained from the three country-specific Amazon Marketplace platforms we consider (Amazon.de; Amazon.co.uk; Amazon.fr). The eight samples include 2,276 products observed for T = 242 days between May 27, 2020, and January 23, 2021 (N = 445,042 daily product-level observations).

Three of the eight samples consist of the relevant best-selling products over the observation period—across all product categories—on each of the three Marketplace websites, respectively (Amazon.de: N = 87,488 observations corresponding to 452 products; Amazon.co.uk: N = 96,316 observations corresponding to 487 products; Amazon.fr: N = 86,678 observations corresponding to 432 products). Additionally, we pool those three samples in a separate data set to increase our tests' statistical power.

These "best-selling samples" only include products with an average sales rank between 1 and 1000. They only include products that exhibit varying buy box sellers in the observation period to ensure that we can identify self-preferencing by measuring the same product over time and assessing the impact of different buy box sellers on visibility.

Beyond those samples, to analyze narrower subcategories that do not necessarily include top-selling products, we collected five additional samples from the German Amazon Marketplace (Amazon.de). One of these samples (N = 93,125 observations corresponding to 483 products) comprises moderately selling products on the website, defined as products with an average sales rank between 10,000 and 11,000. The other four samples correspond, respectively, to four diverse product categories—groceries (N = 19,871 observations corresponding to 98 products), scents (N = 19,454 observations corresponding to 101 products), backpacks (N = 32,787 observations corresponding to 173 products), and batteries (N = 9,323 observations corresponding to 50 products). Each sample contains all relevant products in the respective category with an average sales rank between 1 and 10,000. The diversity of our samples ensures that our findings are not limited to one type of product, category, or country.

Table 4 displays the combined summary statistics for all observations pooled across the three best-selling samples (N = 270,482). In this analysis, the protected attribute for which we estimate the recommendation's self-preferencing, $\hat{\delta}$, is defined as *isAmazonBuyBox*, an indicator variable that takes the value of 1 if Amazon holds the buy box on a focal day and 0 if the buy box holder is a third party.

Amazon maintains a significant presence in our three top-selling samples, securing the buy box for 70.4% of the daily observations. In the remaining 29.6% of observations, third-party sellers supply the product through the buy box. When looking at the entire platform, instead of the best-selling products, the distribution flips: our analysis of data from Keepa.com, in February 2024, shows that Amazon only supplies approximately one-fourth of all products with an active buy box on the German, French, and British marketplaces. Hence, Amazon apparently prioritizes competing for the buy box of best-selling products.

**Table 4:** Study A—Summary Statistics for Pooled Best-Selling Samples from the German, French, and British Marketplaces.

| Variable | Buy Box Seller | Min | q1 | Median | Mean | q3 | Max | SD |
|---|---|---|---|---|---|---|---|---|
| organicVisibility | Third Party | .0 | .1 | .9 | 3.7 | 4.0 | 314.4 | 8.3 |
| | Amazon | .0 | .2 | 1.5 | 4.6 | 5.3 | 309.2 | 9.3 |
| | all | .0 | .1 | 1.3 | 4.3 | 4.9 | 314.4 | 9.0 |
| salesRank | Third Party | 1.0 | 164.0 | 386.0 | 905.0 | 747.0 | 549,998.0 | 7,215.8 |
| | Amazon | 1.0 | 114.0 | 290.0 | 658.0 | 608.0 | 439,948.0 | 5,664.6 |
| | all | 1.0 | 126.0 | 317.0 | 730.8 | 651.0 | 549,998.0 | 6,163.3 |
| price | Third Party | 1.2 | 8.8 | 16.5 | 36.9 | 32.7 | 1,754.0 | 88.0 |
| | Amazon | .7 | 8.8 | 19.3 | 39.0 | 40.0 | 1,849.7 | 74.6 |
| | all | .7 | 8.8 | 18.0 | 38.4 | 38.2 | 1,849.7 | 78.8 |
| countReviews | Third Party | 1.0 | 486.0 | 1,269.0 | 3,449.9 | 3,260.0 | 608,163.0 | 14,614.5 |
| | Amazon | 1.0 | 523.0 | 1,433.0 | 3,681.9 | 3,681.0 | 397,017.0 | 9,966.2 |
| | all | 1.0 | 512.0 | 1,385.0 | 3,613.4 | 3,560.0 | 608,163.0 | 11,535.9 |
| ratingProduct | Third Party | 2.3 | 4.4 | 4.6 | 4.5 | 4.7 | 5.0 | .3 |
| | Amazon | 2.8 | 4.4 | 4.6 | 4.5 | 4.7 | 5.0 | .2 |
| | all | 2.3 | 4.4 | 4.6 | 4.5 | 4.7 | 5.0 | .2 |
| ratingSeller | Third Party | .0 | 91.0 | 94.0 | 92.8 | 97.0 | 100.0 | 6.9 |
| | Amazon | 100.0 | 100.0 | 100.0 | 100.0 | 100.0 | 100.0 | .0 |
| | all | .0 | 99.0 | 100.0 | 98.1 | 100.0 | 100.0 | 4.8 |
| isPrime | Third Party | .0 | .0 | .0 | .4 | 1.0 | 1.0 | .5 |
| | Amazon | .0 | 1.0 | 1.0 | 1.0 | 1.0 | 1.0 | .0 |
| | all | .0 | 1.0 | 1.0 | .8 | 1.0 | 1.0 | .4 |

*Notes: organicVisibility:* non-sponsored visibility of the focal product, as described in equations (1) to (6); *salesRank:* product-specific best-selling rank within the product's broadest category; *price* including shipping fee; *countReviews:* number of reviews for the focal product; *ratingProduct:* the product's rating in stars between 1 and 5. *ratingSeller:* the current buy box seller's rating (between 0 and 100); *isPrime:* indicator equals 1 if the focal product is available for shipment via Prime.
Prices in Pounds converted to Euros with the average 2020 exchange rate (1 Pound = 1.1248 Euros) and N = 270,482 daily product observations.

Table 4 shows that, across all three countries, the mean visibility for observations for which Amazon held the buy box was higher than for observations for which third-party sellers held the buy box. Specifically, products with Amazon holding the buy box, compared to third parties holding the buy box, were, on average, 11.4% more visible in Germany, 21.8% more visible in France, and 38.6% more visible in the United Kingdom. Additionally, across the three best-selling samples, Amazon buy box observations were 5.7% more expensive than observations for which third-party sellers held the buy box. Despite Amazon's higher average price, observations for which Amazon held the buy box, on average, exhibited a 27.3% lower product sales rank compared with observations for which third parties held the buy box, meaning that Amazon-supplied offers sold considerably better than third-party-supplied offers.

The observations in both groups exhibit approximately the same average product rating (4.5 stars out of 5), and observations with Amazon as the buy box seller had 6.7% more reviews than observations with third parties as the buy box seller.

The average seller rating of third parties is 92.8 across our three best-selling samples. Since Amazon lacks a seller rating, we assume a perfect seller rating (100) if Amazon holds the buy box. While this assumption might be reasonable because of Amazon's reputation for high-quality logistics services and customer service, it might bias our coefficient estimates. Hence, as a robustness check, we estimate the model's coefficients without the seller rating and with Amazon seller ratings of 80, 90, and 95. None of these specifications lead to different empirical conclusions. In the following section, we incorporate the variables mentioned above into our model to estimate whether Amazon's search algorithm self-preferences Amazon's over third-party products.

*Results*

Table 5 presents the model estimation results for the eight samples in Models (1) – (8) and the pooled best-selling sample comprising the combined German, French, and UK samples in Model (0). Figure 4 visualizes the self-preferencing estimates $(\exp(\hat{\delta}) - 1) \times 100\%$ for *isAmazonBuyBox* and their 95% confidence intervals.

**Table 5:** Study A—Generalized Linear Model With Log-Link: Estimation Results

| | Dependent variable: organicVisibility | | | | | | | | |
|---|---|---|---|---|---|---|---|---|---|
| Country: | GER, FR, UK | GER | FR | UK | GER | GER | GER | GER | GER |
| Sample type: | best-selling | best-selling | best-selling | best-selling | groceries | batteries | scents | backpacks | medium-selling |
| Model | (0) | (1) | (2) | (3) | (4) | (5) | (6) | (7) | (8) |
| *Variables* | | | | | | | | | |
| isAmazonBuyBox | .016 | .042 | .048** | -.041 | .040 | .213 | .022 | .072 | -.035 |
| | (.029) | (.048) | (.023) | (.055) | (.062) | (.170) | (.026) | (.051) | (.058) |
| ln salesRank | -.120*** | -.100*** | -.100*** | -.140*** | -.009 | -.204** | -.120*** | -.049* | -.137*** |
| | (.010) | (.018) | (.013) | (.017) | (.036) | (.087) | (.017) | (.027) | (.030) |
| ln price | -.368*** | -.212 | -.516*** | -.277*** | -.883*** | -.115 | -.320** | -.300* | -.535*** |
| | (.073) | (.153) | (.098) | (.105) | (.225) | (.409) | (.130) | (.165) | (.153) |
| isPrime | .029 | .029 | .022 | .058 | -.025 | .024 | -.027 | -.172*** | .054 |
| | (.032) | (.065) | (.030) | (.053) | (.075) | (.162) | (.027) | (.054) | (.063) |
| ratingProduct | .104 | .314 | .238 | -.072 | .400 | -1.65** | .227 | -.078 | -.289 |
| | (.161) | (.305) | (.200) | (.206) | (.321) | (.691) | (.275) | (.139) | (.198) |
| ln countReview | .033 | .044 | .075* | -.013 | -.110* | .095 | -.147** | -.021 | -.022 |
| | (.044) | (.052) | (.042) | (.079) | (.063) | (.135) | (.066) | (.068) | (.060) |
| ratingSeller | .000 | -.002 | .000 | .000 | -.003 | -.012 | .002 | .008*** | -.001 |
| | (.001) | (.003) | (.002) | (.001) | (.005) | (.010) | (.002) | (.003) | (.002) |
| *Fixed effects* | | | | | | | | | |
| Product | Yes | Yes | Yes | Yes | Yes | Yes | Yes | Yes | Yes |
| Date | Yes | Yes | Yes | Yes | Yes | Yes | Yes | Yes | Yes |
| No. of products | 1,344 | 452 | 432 | 487 | 98 | 50 | 101 | 173 | 483 |
| Observations | 270,482 | 87,488 | 86,678 | 96,316 | 19,871 | 9,323 | 19,454 | 32,787 | 93,125 |
| Pseudo R² | .713 | .700 | .698 | .749 | .496 | .566 | .444 | .573 | .534 |

*Notes: Two-way (product and date) standard errors in parentheses.*
  *Significance Codes: \*\*\*: .01, \*\*: .05, \*: .1*

**Figure 4:** Study A—Self-Preferencing Point Estimates $\hat{\delta}$ for the Amazon Buy Box Indicator Variable and 95% Confidence Intervals.

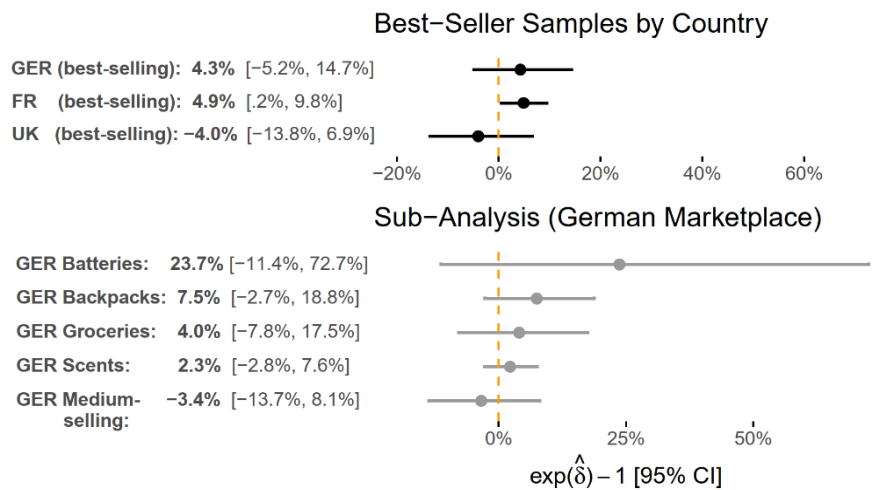

*Notes:* We transform the point estimates from Table 5 by $(exp(\hat{\delta}) - 1) \times 100\%$ to arrive at the exact estimates. Similarly, we compute the exact 95% confidence intervals through $(exp(\hat{\delta} \pm 1.96 \times SE(\hat{\delta})) - 1) \times 100\%$.

For the best-selling samples, the value of the self-preferencing estimate is positive in the German sample and in the French sample, and it is negative in the UK sample (Germany: 4.3%, *p* = 0.390; France: 4.9%, *p* = 0.040; United Kingdom: - 4.0%, *p* = 0.457). Among these three samples, the self-preferencing coefficient is significantly different from zero at the five percent significance level only for the French sample. To decrease the probability of suffering a type-2 error (incorrectly rejecting the null hypothesis) that influences our conclusions, we additionally pool the three best-selling samples and estimate the self-preferencing coefficient. This estimate is also statistically indistinguishable from zero at 1.6% (*p* = 0.556).

For each of the four product-specific samples from the German marketplace, the value of the self-preferencing coefficient estimate is positive, ranging between 2.3% (scents) and 23.7% (batteries). For the medium-selling sample, the coefficient is slightly negative (-3.4%). However, all estimates are statistically indistinguishable from zero at *p* > 0.1.

The positive and significant self-preferencing coefficient for the best-selling sample from France may suggest self-preferenced search results. Nevertheless, since only one out of nine estimates is significantly positive, this analysis provides almost no evidence that Amazon's search results prefer products for which Amazon occupies the buy box.

**Study B: Testing for Self-Preferencing Through Recommendations Among Private-Label Products**

In Study A, we scrutinized the effect of the *within*-product variation of the buy box seller on product visibility. Study B augments the analysis with a *between*-product comparison of products exclusively sold by either Amazon or third parties. As alluded to above, a seller—including Amazon—can avoid competing for a product's buy box if the seller is the sole supplier of that product. This setting is the case for certain private-label products; for example, Amazon is the only seller on the platform offering products from its Amazon Basics brand. In this case, the products still compete for search engine visibility against other similar products from third-party sellers. Hence, in Study B, we examine whether, after controlling for unprotected attributes that might influence search visibility, Amazon's private-label products (here: Amazon Basics products) are more visible in Amazon Marketplace search results compared with similar third-party private-label substitutes.

*Sample Selection and Descriptive Statistics*

In constructing the sample for this study, we sought to create comparison groups of Amazon Basics products and their comparable product substitutes exclusively supplied by third-party sellers. As in Study A, we collected data from the German (Amazon.de), British (.co.uk), and French (.fr) marketplaces.

We collected observations for all Amazon Basics products on each of the three Amazon Marketplaces of products first listed before 2020. That filter ensures that we only compare reasonably well-established products within our observation period (May 27, 2020, to January 22, 2021). We dropped all products with an average sales rank greater than 50,000 to exclude infrequently sold products. Further, we excluded all products unavailable for more than 50% of the observation period.

Next, for each Amazon Basics product in the dataset, we identified product substitutes that were also third-party private-label products. To this end, we relied on the following procedure, elaborated by Clara (2023):

Amazon categorizes products in a category tree, starting with approximately 30 broad root categories and becoming narrower with every node. Products that share the narrowest category are very similar and serve as close substitutes. For example, the category tree for the product *Amazon Basics Waterproof Snow Gloves* consists of the following six levels: *Sports & Outdoors > Winter Sports > Skiing > Clothing > Men > Men's Ski Gloves*. In this category tree, *Sports & Outdoors* is the broadest root category, and *Men's Ski Gloves* represents a very narrow subgroup of close substitutes. For every Amazon Basics product, we obtained all third-party-supplied products from the product's narrowest subcategory that fulfilled the same selection criteria as the Amazon Basics products (first listing date, sales rank, and availability) and additionally had a single buy box seller that is not Amazon. Restricting the set to products with a single third-party buy box seller ensures that we only compare (Amazon vs. third party) private label products to each other, increasing those products' comparability. Next, we created comparison groups consisting of a single Amazon Basics product and up to five private-label substitutes. If a particular Amazon Basics product had more than five substitutes, we randomly sampled five substitutes from the larger set.

Our three samples collectively include 592,742 daily product-level observations, corresponding to 749 Amazon Basics products and 2,478 third-party substitute products observed on T = 241 days from May 27, 2020, to January 22, 2021. Split by country, Study B's sample includes 291 Amazon Basics products and their close substitutes (i.e., comparison groups) with 217,446 daily observations for Amazon.de, 238 comparison groups with 193,248 daily observations for Amazon.fr, and 223 comparison groups with 182,048 daily observations for Amazon.co.uk.

Table 6 displays detailed summary statistics for the observations. In this analysis, the variable for which we estimate the self-preferencing coefficient $\delta$ is defined as *isAmazonBasics*, an indicator variable that takes the value of 1 if the product is an Amazon Basics product, and 0 if it is a third-party private-label product.

Table 6 shows that in contrast to Study A, where products currently supplied by Amazon were more visible than third-party products, Amazon Basics products' mean organic search visibility is slightly below the mean visibility of third parties' comparable products (- 9.6% over all three samples). Furthermore, Amazon Basics products, on average, exhibit a 33.7% lower sales rank (meaning they sold better) and are only marginally cheaper (1.0%) than their substitutes at an average price of 21.1 € vs. 20.9 €. While the average ratings of the two product groups are almost identical at 4.5 and 4.4 stars, Amazon Basics products, on average, have 4.5 times as many reviews compared to their close third-party competitors (4,919.4 vs. 1,102.5).

**Table 6:** Study B—Summary Statistics by Amazon Basics vs. Third-Party Private-Label Products.

| Variable | Product | Min | q1 | Median | Mean | q3 | Max | SD |
|---|---|---|---|---|---|---|---|---|
| organicVisibility | Amazon Basics | .0 | .1 | .3 | 1.4 | 1.2 | 189.6 | 4.8 |
| | Third-Party Private Label | .0 | .1 | .4 | 1.6 | 1.5 | 106.7 | 3.6 |
| | all | .0 | .1 | .4 | 1.5 | 1.4 | 189.6 | 3.9 |
| salesRank | Amazon Basics | 1.0 | 462.0 | 1,607.0 | 3,255.7 | 4,274.0 | 214,948.0 | 5,100.1 |
| | Third-Party Private Label | 1.0 | 1,185.0 | 2,925.0 | 4,910.8 | 5,948.0 | 1,269,875.0 | 20,219.6 |
| | all | 1.0 | 955.0 | 2,626.0 | 4,521.1 | 5,600.0 | 1,269,875.0 | 17,865.3 |
| price | Amazon Basics | 2.7 | 9.6 | 14.7 | 20.9 | 23.2 | 404.5 | 20.5 |
| | Third-Party Private Label | 0.9 | 8.5 | 14.2 | 21.1 | 24.0 | 489.0 | 27.8 |
| | all | .9 | 8.9 | 14.4 | 21.1 | 23.6 | 489.0 | 26.2 |
| countReviews | Amazon Basics | 1.0 | 429.0 | 1,308.0 | 4,919.4 | 4,481.0 | 301,833.0 | 13,609.5 |
| | Third-Party Private Label | 1.0 | 137.0 | 373.0 | 1,102.5 | 1,013.0 | 48,950.0 | 2,379.5 |
| | all | 1.0 | 164.0 | 485.0 | 2,001.2 | 1,462.0 | 301,833.0 | 7,110.5 |
| ratingProduct | Amazon Basics | 2.5 | 4.4 | 4.5 | 4.5 | 4.6 | 5.0 | .2 |
| | Third-Party Private Label | 1.0 | 4.3 | 4.5 | 4.4 | 4.6 | 5.0 | .3 |
| | all | 1.0 | 4.3 | 4.5 | 4.5 | 4.6 | 5.0 | .3 |
| ratingSeller | Amazon Basics | 100.0 | 100.0 | 100.0 | 100.0 | 100.0 | 100.0 | .0 |
| | Third-Party Private Label | .0 | 93.0 | 97.0 | 94.7 | 98.0 | 100.0 | 6.3 |
| | all | .0 | 95.0 | 98.0 | 96.1 | 100.0 | 100.0 | 5.9 |
| isPrime | Amazon Basics | 1.0 | 1.0 | 1.0 | 1.0 | 1.0 | 1.0 | .0 |
| | Third-Party Private Label | .0 | .0 | 1.0 | .7 | 1.0 | 1.0 | .5 |
| | all | .0 | .0 | 1.0 | .7 | 1.0 | 1.0 | .4 |

*Notes: organicVisibility:* non-sponsored visibility of the focal product, as described in equations (1) to (6); *salesRank:* product-specific best-selling rank within the product's broadest category; *price* including shipping cost; *countReviews:* number of reviews for the focal product; *ratingProduct:* the product's rating in stars between 1 and 5. *ratingSeller:* the current buy box seller's rating (between 0 and 100); *isPrime*: indicator equals 1 if the focal product is available for shipment via Prime. Prices in Pounds converted to Euros with the average 2020 exchange rate (1 Pound = 1.1248 Euros) and N = 270,482 daily product observations.

*Results*

To determine whether Amazon's search results self-preference against third-party private-label products in favor of Amazon Basics products, we conducted a regression analysis similar to that used in Study A. The regressions differ, however, in the comparison unit; that is, whereas the regression in Study A compared visibility within products, where the variable of interest was *isAmazonBuyBox*, here we compare visibility between Amazon Basics and third-party products (*isAmazonBasics*).

Because we are comparing between products, it is not appropriate to control for individual product fixed effects (as in Study A), as doing so would remove any heterogeneous effect of Amazon Basics products on visibility. Accordingly, we replace product-fixed effects with comparison-group-fixed effects that cluster one Amazon Basics product with up to five comparable private-label substitutes. In this case, the comparison-group-fixed effects absorb the heterogeneity across comparison groups but not across individual products *within* comparison groups.

Table 7 presents the detailed estimation results. We visualize and compare the point estimates and associated 95% confidence intervals of the self-preferencing estimate $\hat{\delta}$ for each of the three country samples in Figure 5. In contrast to our results for Study A, we estimate a substantial and statistically highly significant *negative* Amazon effect in all three (German, British, and French) samples. These findings imply that Amazon Basics products are less visible than justified in the organic search results.

Estimates of the coefficient $\delta$ range from $(\exp(\hat{\delta}_{FR}) - 1) \times 100\% = -46.0\%$ for the French sample to –48.2% and –49.5% for the British and German marketplaces, respectively. Additionally, the coefficient estimates' 95% confidence intervals lie substantially within the negative domain in all three cases. These results suggest that, in all three marketplaces, it is highly unlikely that Amazon's organic search results suffer from self-preferencing against third-party private-label products in favor of Amazon Basics products.

**Table 7:** Study B—Generalized Linear Model With Log-Link: Estimation Results.

|  | Dependent variable: organicVisibility | | |
|---|---|---|---|
| Country: | GER | FR | UK |
| Model: | (1) | (2) | (3) |
| *Variables* | | | |
| isAmazonBasics | -.682*** | -.616*** | -.658*** |
|  | (.123) | (.146) | (.174) |
| ln salesRank | -.304*** | -.356*** | -.360*** |
|  | (.031) | (.022) | (.041) |
| ln price | .203** | .202*** | -.225 |
|  | (.089) | (.073) | (.144) |
| isPrime | .075 | .252 | -.072 |
|  | (.210) | (.174) | (.151) |
| ratingProduct | -.427* | -.006 | .299 |
|  | (.238) | (.155) | (.318) |
| ln countReview | .350*** | .114*** | .167*** |
|  | (.033) | (.029) | (.032) |
| ratingSeller | -.010 | -.008* | .002 |
|  | (.009) | (.004) | (.006) |
| *Fixed effects* | | | |
| Comparison group | Yes | Yes | Yes |
| Date | Yes | Yes | Yes |
| No. of comparison groups | 291 | 238 | 223 |
| Observations | 217,446 | 193,248 | 182,048 |
| Pseudo R² | .433 | .417 | .433 |

*Notes: Two-way (comparison group and date) standard errors in parentheses.*
*Significance Codes: \*\*\*: .01, \*\*: .05, \*: .1*

**Figure 5:** Study B— Self-Preferencing Point Estimates $\hat{\delta}$ for the Amazon Basics Indicator Variable and 95% Confidence Intervals.

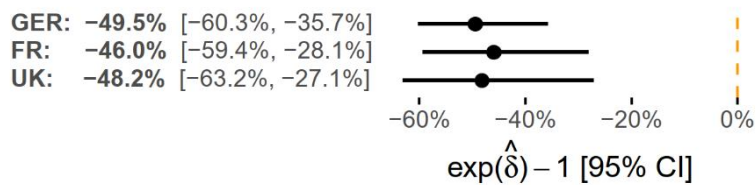

*Notes:* We transform the point estimates from Table 7 by $(exp(\hat{\delta}) - 1) \times 100\%$ to arrive at the exact estimates. Similarly, we compute the exact confidence intervals through $(exp(\hat{\delta} \pm 1.96 \times SE(\hat{\delta})) - 1) \times 100\%$.

The validity of Study B rests upon the assumption that all the Amazon Basics and third-party private-label products in a comparison group are indeed comparable. Stated differently, if we could estimate product-fixed effects in addition to the observed attributes (e.g., product reviews, seller reviews, price), their values should be equal for all products in a category. Unfortunately, we cannot do so while also estimating the coefficient for *isAmazonBasics*. What we can do, however, is vary the set of comparison groups to test if the results hold when excluding those comparison groups that could be suspected to differ in unobserved attributes. We do so in Web Appendix D, where we exclude comparison groups for which the Amazon Basic product's mean organic visibility differs to a varying degree from the mean of the respective third-party competitors' products. The results remain consistent for all possible thresholds, providing reassurance either that the products in the comparison groups are reasonably similar or that any potential difference does not affect the study's conclusions.

### Comparison of Results Between Study A and B

In Study A, we compared the visibility of offers characterized by identical products but different buy box sellers. In contrast, Study B looked at the visibility of comparable private-label products and compared one Amazon Basics product with up to five closely competing products. Counterintuitively, Amazon Basics products received far lower visibility and, thus, fewer recommendations than "deserved," according to our self-preferencing test. The substantially negative values for the self-preferencing estimates from Study B differ from the results in Study A, where we could not identify a statistically significant difference in most samples. There are several potential reasons for this divergence. While there could be data or econometric limitations, which we carefully discuss and address in the following section, there are plausible business-related explanations for these findings.

One possible explanation is that Amazon Basics products are clearly associated with the platform provider Amazon. In contrast, the current buy box seller is a subtle attribute that consumers and regulators cannot easily compare across products and search results. If Amazon purposely wanted to avoid the impression of favoring its own products in the search results, then it could, on the one hand, assign a lower visibility to its Amazon Basics products such that casual observers do not get the impression that those products are favored. On the other hand, such "cautionary" lower visibility would be less necessary for products that Amazon currently supplies via the buy box since consumers and regulators cannot easily identify and compare the current buy box seller through the search results but only observe this attribute after clicking on each listing.

Additionally, for Amazon Basics products, the platform provider Amazon might compensate for the apparent lack of *organic* visibility by higher *sponsored* visibility. In Study B's samples, Amazon Basic products, on average, display 2.9 times higher sponsored visibility than comparable third-party products. Eventually, this difference might compensate for the lower organic visibility, since Amazon Basics products are still very visible, though labeled as "sponsored" results. This practice might evade criticism from researchers and regulators since Amazon labels them as non-organic. However, we stress that such a practice is merely a conjecture, and further research would be required to corroborate it. Evaluating whether this practice is legitimate is beyond the scope of this study and could require data on prices for sponsored search results and internal accounting mechanisms. In the future, researchers or regulators could examine if Amazon retail "pays" a market price to the platform provider Amazon for sponsored search results.

### Robustness of the Results and Limitations

*Robustness to Omitted Variable Bias*

Our approach to detecting self-preferencing requires that we account for all unprotected attributes that influence the recommendation and correlate with the protected attribute. In our two studies on the Amazon marketplace, we used two main approaches to ensure that we would not miss such a variable. First, we matched Amazon's own disclosure of attributes influencing the search results—disclosed in response to the European Union's platform-to-business regulation (Amazon 2022; European Parliament 2019)—to our observable attributes. Second, we used a consumer survey to identify product attributes influencing consumer choice and made sure to consider these attributes in our analysis.

Nevertheless, these efforts are not foolproof, meaning there is still a risk that we might have missed a relevant unprotected attribute. Therefore, we implement two additional analyses to ensure that omitting such a variable does not change our conclusions. First, in Web Appendix A, we implement the alternative OB test for self-preferencing (Reimers and Waldfogel 2023), which relies on different assumptions and is less sensitive to omitted variable bias. The results of this alternative test provide no evidence of self-preferencing, thus supporting our main analysis.

Second, we use Web Appendix C to assess under which conditions an omitted variable could change our conclusions from "finding no evidence for self-preferencing" to "finding evidence for self-preferencing." We show that only an omitted variable that represents an "Amazon disadvantage," i.e., a variable that is worse for Amazon's products and positively influences visibility, could threaten our conclusion because it would positively bias our self-preferencing estimate. Because of Amazon's reputation for high-quality customer service and logistics, the existence of such an omitted variable is unlikely. In contrast, omitting a variable representing an "Amazon advantage" would bias our self-preferencing estimate negatively. Because we estimate self-preferencing coefficients close to zero (Study A) or negative ones (Study B), an additional negative bias would not change our conclusion of finding little evidence for self-preferencing.

*Robustness to Measurement Error Causing the Null Results*

Since we combine two data sources that measure their variables at potentially slightly different times of the day, our data might suffer from measurement error. The likelihood of measurement error influencing the conclusions of Study B is small due to the relatively large negative estimates. Nonetheless, measurement error could be problematic in Study A, where we did not identify a statistically significant coefficient of the *isAmazonBuyBox* indicator. Hence, we could fail to reject the null hypothesis even though there would be a true effect ("type-2 error"). Through a sensitivity analysis in Web Appendix B, we provide empirical evidence of why potential measurement error is unlikely to influence our conclusions. Additionally, the results still hold when pooling the distinct samples to increase statistical power further.

**Discussion**

Amid regulatory and litigation efforts to prevent self-preferencing on digital platforms, this research sought to address the urgent need for a robust toolkit for identifying self-preferencing when it occurs, to the benefit of all stakeholders involved. To this end, we conceptualized two tests, and implemented them to test for self-preferencing in Amazon's search results. To better understand the implications of our endeavor, we prefaced our analyses with a survey of consumer perceptions regarding self-preferencing on Amazon. This final section summarizes the key findings, concludes what we can learn from them, and discusses the findings' implications for stakeholders on digital platforms.

*Consumers Expect Self-Preferencing, Impacting Trust but not Future Purchases*

Our consumer survey showed that consumers overwhelmingly expect Amazon to engage in self-preferencing through its search results. In addition, consumers stated that self-preferencing diminishes their trust in the platform. However, only a few consumers reported that they would change their future purchasing behavior if an investigation found self-preferenced search results.

Notably, consumers' expectations regarding self-preferencing are misaligned with the results of our empirical analysis, which revealed almost no evidence for self-preferencing. This misalignment suggests that the platform provider unnecessarily loses consumer trust. Hence, platforms could benefit from better informing consumers regarding the validity of their recommendation systems, e.g., by letting independent third parties audit their recommendations for self-preferencing, using methodologies such as the one developed herein.

The survey findings also indicate that even if Amazon did engage in self-preferencing, it would face minimal "punishment" from consumers. This finding could imply either that consumers perceive the utility of using Amazon as outweighing the perceived costs that self-preferencing inflicts, or that Amazon has so much market power that consumers could not abandon the platform, even if they wanted to because of self-preferencing. This finding highlights the importance of regulations that ban self-preferencing—whose efficacy is contingent on tools, such as ours, for the robust assessment of such behavior. Indeed, given that consumers might feel unable to adapt their behavior on platforms in response to self-preferencing, it might be the case that regulations are one of the few means to protect consumers and third-party actors from the loss of surplus that self-preferencing could inflict.

*Almost no Evidence for Self-Preferencing in Amazon's Search Results*

In our main analysis, we modeled the recommendation (visibility) as a function of a protected attribute (*isAmazon*) and unprotected attributes, such as prices, that can legitimately influence the recommendation. We applied this COO test for self-preferencing via search results in two competitive settings on the Amazon marketplace. In both settings, we found almost no evidence for self-preferencing. Across Studies A and B, we identified weakly self-preferenced search results in only one out of eleven samples from three countries. In all other settings, we either found an effect that was statistically insignificantly different from zero (Study A) or revealed statistically significant negative self-preferencing coefficients, meaning that third-party products are more prominent in the search results than their attributes would justify (Study B).

These results suggest that, perhaps surprisingly, third-party sellers might even have an advantage through the search results if they compete with Amazon Basics products or not have a disadvantage if they compete for sales of the same product with Amazon through the buy box. One potential explanation for this result might be that Amazon still has a strong incentive to sell third-party products via its platform: compared to bearing the full risk of manufacturing, supplying, and retailing their (private-label) products, Amazon faces less risk when only serving as the intermediary for sales of third-party products, while still earning a sizable commission. Hence, the additional profits through higher integration of a product's value chain might not outweigh the additional risks. Moreover, it is possible that Amazon might self-preference through other practices outside this study's scope, as we elaborate further below.

*Empirical Challenges and Solutions when Implementing Self-Preferencing Tests*

Throughout the manuscript, we have outlined the relevant assumptions for our main COO test to produce unbiased self-preferencing estimates. The key assumption is that we do not omit any relevant variable that influences the recommendation and correlates with the protected attribute. While we provide a multilayered approach that uses regulatory disclosure and a consumer survey to ensure we do not omit any such variable, we cannot fully rule out the possibility that such omission occurred. Hence, for robustness, we also implemented the alternative OB test for self-preferencing that relies on a different set of assumptions and does not need to control for all attributes influencing the recommendation because it models the outcome (in our case, sales ranks) as a function of the recommendation (see Web Appendix A). While the OB test comes with its own challenges, estimating self-preferencing with both tests can help assess the robustness of conclusions to violations of a test's assumptions.

In our case, neither test finds much evidence for self-preferencing. Specifically, both tests arrive at almost the same results in Study A (self-preferencing estimates indistinguishable from zero). In Study B, whereas the COO test produced a negative self-preferencing estimate in all three samples, the OB test produced a self-preferencing estimate statistically insignificantly different from zero in two out of three samples; in the remaining sample, the self-preferencing estimate was negative. Hence, both tests arrive at the same conclusion of not finding evidence for self-preferencing. This robustness of the results reassures us that our initial result did not suffer from omitted variable bias that would change our conclusion.

Overall, these considerations show that analysts conducting self-preferencing tests need to ensure that their conclusions do not change due to limitations of the tests or omitted variables. Using the two suggested tests and verifying that they yield the same conclusions helps establish confidence in the results.

*Scope of our Findings within the Self-Preferencing Debate*

We emphasize that our empirical illustration applies our approach to two forms of competition, albeit on a single platform and with a single type of recommendation (organic search engine visibility). Even though we find almost no evidence for self-preferencing through organic search engine visibility on the Amazon Marketplace, we cannot conclude that self-preferencing never occurs. Amazon has several other opportunities to self-preference, which are harder to evaluate as an outside researcher; for example, Amazon might preference its own products through recommendation formats such as "frequently bought together" labels or product features on Amazon's home page.

For a more complete assessment, future studies might examine, for example, whether Amazon self-preferences by labeling certain offers as "Amazon's Choice." To this end, analysts could replace the dependent variable in our model with an appropriate indicator variable (e.g., *isAmazonChoice*). In particular, our approach could estimate whether Amazon's own products are more or less likely than third-party products to be labeled "Amazon's Choice" after accounting for unprotected attributes that influence such a recommendation. Researchers or regulators could also apply our tests for self-preferencing to platforms outside Amazon Marketplace that supply their own first-party offers, provide search engine results or other recommendations, and compete with other third-party offers.

*Applicability of Our Approach to Other Platforms and Recommendation Types*

While our empirical section focused on two competitive settings on the Amazon marketplace, our conceptual framework containing the two self-preferencing tests also applies to other platforms recommending offers to consumers. The applicability of those tests depends on data availability. Our study used data from Sistrix to measure search visibility on the Amazon Marketplace and Keepa for all other product data. While commercial data providers provide a straightforward means of gathering relevant data from platforms such as Amazon and Google, an analyst could deploy any web crawler that repeatedly tracks how prominently the platform recommends an offer.

While this manuscript has considered the conceptual framework and the two associated tests as tests for *self*-preferencing, analysts can easily repurpose the COO and OB tests to identify *preferencing* of different sorts. For example, analysts can replace the protected attribute *isPlatform* to estimate whether particular offer groups are recommended differently based on other protected attributes, such as the personal attributes (e.g., race or gender) of humans involved in the offer supply.

Another possibility is to replace the protected attribute *isPlatform* with an unprotected offer attribute, as a means of gaining insight into which attributes influence the recommendation. Hence, beyond testing for self-preferencing, the COO test could suggest how platform actors should position their offers regarding pricing, quality, or other characteristics to receive favorable recommendations from the platform provider.

# References

Agarwal, Ashis, Kartik Hosanagar, and Michael D. Smith (2011), "Location, Location, Location: An Analysis of Profitability of Position in Online Advertising Markets," *Journal of Marketing Research*, 48 (6), 1057–73.

Aguiar, Luis and Joel Waldfogel (2021), "Platforms, Power, and Promotion: Evidence from Spotify Playlists," *The Journal of Industrial Economics*, 69 (3), 653–91.

Aguiar, Luis, Joel Waldfogel, and Sarah Waldfogel (2021), "Playlisting favorites: Measuring platform bias in the music industry," *International Journal of Industrial Organization*, 78, 102765.

Amazon (2022), "Finding Products in the Amazon Store," (accessed November 3, 2022), https://web.archive.org/web/20221103133008/https://www.amazon.de/gp/help/customer/display.html?nodeId=504960&language=en_GB.

Amazon Seller Central (2022), "Where and What to Sell with Amazon Global Selling," (accessed March 23, 2022), https://sellercentral.amazon.com/gp/help/external/G201468340.

Barach, Moshe A., Joseph M. Golden, and John J. Horton (2020), "Steering in Online Markets: The Role of Platform Incentives and Credibility," *Management Science*, 66 (9), 4047–70.

Beus, Johannes (2015), "Click probabilities in the Google SERPs," *SISTRIX*, (accessed January 22, 2024), https://www.sistrix.com/blog/click-probabilities-in-the-google-serps/.

Beus, Johannes (2020), "Why (Almost) Everything You Knew About Google CTR is no Longer Valid," *SISTRIX*, (accessed April 13, 2022), https://www.sistrix.com/blog/why-almost-everything-you-knew-about-google-ctr-is-no-longer-valid/.

Cicilline, David N. (2021), "American Innovation and Choice Online Act," *H.R.3816, 117th Congress*, https://www.congress.gov/117/bills/hr3816/BILLS-117hr3816rh.xml.

Clara, Nuno (2023), "Demand Elasticities, Nominal Rigidities and Asset Prices," *working paper*.

Constantinides, Panos, Ola Henfridsson, and Geoffrey G. Parker (2018), "Platforms and Infrastructures in the Digital Age," *Information Systems Research*, 29 (2), 381–400.

Cure, Morgane, Matthias Hunold, Reinhold Kesler, Ulrich Laitenberger, and Thomas Larrieu (2022), "Vertical Integration of Platforms and Product Prominence," *Quantitative Marketing and Economics*, 20, 353–95.

Drèze, Xavier and Fred Zufryden (2004), "Measurement of Online Visibility and its Impact on Internet Traffic," *Journal of Interactive Marketing*, 18 (1), 20–37.

Dubé, Jean-Pierre (2024), "Amazon Private Brands: Self-Preferencing vs Traditional Retailing," *The Antitrust Journal*, 86 (1).

European Parliament (2019), "Regulation (EU) 2019/1150 of the European Parliament and of the Council of 20 June 2019 on promoting fairness and transparency for business users of online intermediation services," (accessed June 28, 2022), https://eur-lex.europa.eu/legal-content/EN/TXT/HTML/?uri=CELEX:32019R1150.

European Parliament (2022), "Regulation (EU) 2022/1925 of the European Parliament and of the Council of 14 September 2022 on contestable and fair markets in the digital sector and amending Directives (EU) 2019/1937 and (EU) 2020/1828 (Digital Markets Act)," (accessed November 10, 2022), https://eur-lex.europa.eu/eli/reg/2022/1925/oj.

Farronato, Chiara, Andrey Fradkin, and Alexander MacKay (2023), "Self-Preferencing at Amazon: Evidence from Search Rankings," *AEA Papers and Proceedings*, 113, 239–43.

Federal Trade Commission (2023), "FTC Sues Amazon for Illegally Maintaining Monopoly Power," (accessed October 11, 2023), https://www.ftc.gov/news-events/news/press-releases/2023/09/ftc-sues-amazon-illegally-maintaining-monopoly-power.

Feng, Juan, Hemant K. Bhargava, and David M. Pennock (2007), "Implementing Sponsored Search in Web Search Engines: Computational Evaluation of Alternative Mechanisms," *INFORMS Journal on Computing*, 19 (1), 137–48.

Ghose, Anindya, Panagiotis G. Ipeirotis, and Beibei Li (2014), "Examining the impact of ranking on consumer behavior and search engine revenue," *Management Science*, 60 (7), 1632–54.

Ghose, Anindya and Sha Yang (2009), "An Empirical Analysis of Search Engine Advertising: Sponsored Search in Electronic Markets," *Management Science*, 55 (10), 1605–22.

Goldfarb, Avi and Catherine Tucker (2019), "Digital Economics," *Journal of Economic Literature*, 57 (1), 3–43.

Hunold, Matthias, Reinhold Kesler, and Ulrich Laitenberger (2020), "Rankings of Online Travel Agents, Channel Pricing, and Consumer Protection," *Marketing Science*, 39 (1), 92–116.

Jeffries, Adrianne and Leon Yin (2021), "Amazon Puts Its Own 'Brands' First Above Better-Rated Products," (accessed February 8, 2023),

https://themarkup.org/amazons-advantage/2021/10/14/amazon-puts-its-own-brands-first-above-better-rated-products.

Khan, Lina (2019), "The Separation of Platforms and Commerce," *Columbia Law Review*, 119 (4), 973–1098.

Lai, Guoming, Huihui Liu, Wenqiang Xiao, and Xinyi Zhao (2022), "'Fulfilled by Amazon': A Strategic Perspective of Competition at the e-Commerce Platform," *Manufacturing & Service Operations Management*, 24 (3), 1406–20.

Lanxner, Eyal (2019), "How to Master and Win the Amazon Buy Box," *The BigCommerce Blog*, (accessed January 14, 2021), https://www.bigcommerce.co.uk/blog/win-amazon-buy-box/.

Long, Fei and Wilfred Amaldoss (2024), "Self-Preferencing in E-commerce Marketplaces: The Role of Sponsored Advertising and Private Labels," *Marketing Science*, forthcoming.

Mattioli, Dana (2019), "Amazon Changed Search Algorithm in Ways That Boost Its Own Products," *Wall Street Journal*, (accessed December 5, 2020), https://www.wsj.com/articles/amazon-changed-search-algorithm-in-ways-that-boost-its-own-products-11568645345.

Pagani, Margherita (2013), "Digital Business Strategy and Value Creation: Framing the Dynamic Cycle of Control Points," *MIS Quarterly*, 37, 617–32.

Raval, Devesh (2022), "Steering in One Click: Platform Self-Preferencing in the Amazon Buy Box," *working paper*.

Reimers, Imke and Joel Waldfogel (2023), "A Framework for Detection, Measurement, and Welfare Analysis of Platform Bias," NBER Working Paper 31766.

Reiner, Jochen, Oliver J. Rutz, and Bernd Skiera (2023), "Investigating a Retail Platform's Recommendation of a Seller – Measuring the Effect of Amazon's Buy Box," *working paper*.

Sistrix (2020), "SISTRIX Visibility Index for Amazon," *Sistrix Documentation*, (accessed December 10, 2020), https://www.sistrix.com/amazon/amazon-visibilityindex.

Sistrix (2021), "SISTRIX Visibility Index — Explanation, Background and Calculation," *Sistrix Documentation*, https://www.sistrix.com/support/sistrix-visibility-index-explanation-background-and-calculation/.

Skiera, Bernd and Nadia Abou Nabout (2013), "PROSAD: A Bidding Decision Support System for Profit Optimizing Search Engine Advertising," *Marketing Science*, 32 (2), 213–20.

Statista (2021), "Leading Global Online Marketplaces 2020, by GMV," *Statista*, (accessed March 23, 2022), https://www.statista.com/statistics/885354/top-global-online-marketplaces-by-gmv/.

de Streel, Alexandre, Jacques Crémer, Paul Heidhues, David Dinielli, Gene Kimmelman, Giorgio Monti, Rupprecht Podszun, Monika Schnitzer, and Fiona M. Scott Morton (2022), "Enforcing the Digital Markets Act: Institutional Choices, Compliance, and Antitrust," *working paper*, https://papers.ssrn.com/abstract=4314848.

Ursu, Raluca M. (2018), "The Power of Rankings: Quantifying the Effect of Rankings on Online Consumer Search and Purchase Decisions," *Marketing Science*, 39 (4), 507–684.

Yin, Leon and Adrianne Jeffries (2021), "How We Analyzed Amazon's Treatment of Its 'Brands' in Search Results – The Markup," (accessed August 29, 2023), https://themarkup.org/amazons-advantage/2021/10/14/how-we-analyzed-amazons-treatment-of-its-brands-in-search-results.

Zeibak, Leanna (2020), "How to Win the Amazon Buy Box in 2021," *Tinuiti Blog*, (accessed January 14, 2021), https://tinuiti.com/blog/amazon/win-amazon-buy-box/.

# Web Appendix A: Robustness Check—Comparison of Conditioning-on-Observables vs. Outcome-Based Test for Biased Recommendations

*Outcome-Based Test: Alternative Test not Relying on Capturing Visibility-Independent Offer Utility*

While the COO test directly estimates self-preferencing in the recommendation, controlling for all attributes that affect consumer choice might not always be possible. An alternative approach not subject to this concern (but with other shortcomings, as elaborated below) is the Outcome-based (OB) test for bias in recommendations (see Reimers and Waldfogel (2023)).

According to the OB test, a platform's recommendations are unbiased if the analyst observes two products with equal visibility that yield an equal outcome—in our case, equal sales on the Amazon Marketplace. Hence, this alternative approach compares offers' success *conditional* on their visibility. Suppose we observe two products—one supplied by Amazon and one supplied by a third party. In that case, the OB test indicates biased visibility in favor of Amazon if Amazon sells less than the third party (and vice versa). This outcome would show that consumers choose the Amazon product less, even though the platform recommends it equally. Hence, the recommendation does not fully reflect consumers' utility.

Accordingly, we model the outcome, in our case *salesRank*, as a function of the platform indicator, the natural logarithm of organic visibility, and product- and day-fixed effects:

$$E[\text{salesRank}_{it}| \cdot ] = \exp\left(\delta_{OB}\text{isAmazon}_{it} + \gamma_1 \ln \widetilde{VI}_{it} + \alpha_i + \gamma_t\right). \quad (8)$$

The parameter $\hat{\delta}_{OB}$ indicates the bias in organic visibility towards Amazon. Because a lower value for *salesRank* signals higher sales, $\hat{\delta}_{OB} > 0$ implies biased visibility in favor of Amazon, consistently with the interpretation of the bias estimate from the COO test.

The advantage of this test is that we can quantify self-preferenced recommendations, i.e., recommendations that position Amazon's products higher than where they "should" be based on consumers' utility from those products, without having to measure this "visibility-independent utility." However, for this approach to yield unbiased self-preferencing estimates, the dependent variable (i.e., the outcome) must directly result from the recommendation without confounding effects of an omitted variable that correlates with both the outcome and the platform indicator.

Hence, the OB test relieves us from the threat that we omit a variable explaining consumers' mean utility, but in turn, introduces the requirement that we need to control for all attributes that correlate both with a product's success—its *salesRank*—and the platform indicator. Thus, Reimers and Waldfogel (2023) include *Price* in their basic regression equation (8). However, estimating such a regression equation is also likely to suffer from omitted variable bias. In our case, advertising for a product through the search engine (i.e., sponsored visibility) should influence the product's *salesRank*, and correlate substantially with the platform indicator. Hence, we have to add a product's sponsored visibility and its price to equation (8) to estimate

$$E[salesRank_{it}| \cdot ] = \exp\left(\delta_{OB}\ isAmazon_{it} + \gamma_1 \ln \widetilde{VI}_{organic,it} + \gamma_2 \ln \widetilde{VI}_{sponsored,it} + \gamma_3 \ln price_{it} + \alpha_i + \gamma_t\right). \quad (9)$$

*Results Comparison: Outcome-Based vs. Conditioning-on-Observables Test*

Analogous to our main analysis, we lag all independent variables in this equation by one day to avoid measurement of the dependent variable before a change in the *isAmazon* variable occurs. We now contrast the alternative OB test estimates with the main analyses' COO test for each of the two studies.

*Study A*

Figure 6 shows that all 95% confidence intervals include zero for the OB test's estimates. Hence, interpreting the estimates from the alternative OB test supports the main result that we fail to identify a positive bias in favor of Amazon. Of the eight samples, we cannot statistically distinguish the estimates from zero in every single sample through the OB test, in comparison to a single statistically significant (p < 0.05) positive estimate in the main COO test.

Because the OB test (in contrast to the COO test) does not rely on accounting for products' visibility-independent mean utility, and both tests support an almost identical conclusion, we regard this robustness check as further evidence that omitted variable bias does not affect the conclusions derived from Study A.

**Figure 6:** Study A—Comparison of Bias Estimates for Conditioning-on-Observables vs. Outcome-Based Test

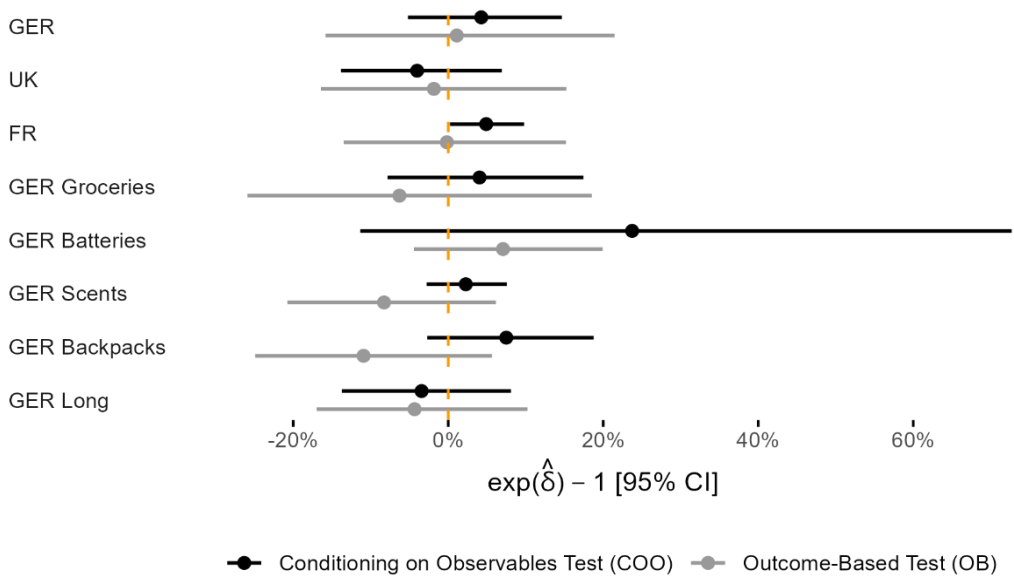

*Study B*

Figure 7 repeats this robustness check for Study B and contrasts the results of the COO and OB tests. For Study B, we estimate no positive effect, i.e., no evidence for self-preferencing in all three samples. While the COO test from the main analysis found consistent negative estimates in all three samples, the OB test replicates this result in the French sample but not in the two other samples (Germany and the UK; their estimates are statistically indistinguishable from zero at $p < 0.05$). Thus, both tests in all three samples are consistent in finding no evidence of self-preferencing of Amazon Basics products.

**Figure 7:** Study B—Comparison of Bias Estimates for Conditioning-on-Observables vs. Outcome-Based Test

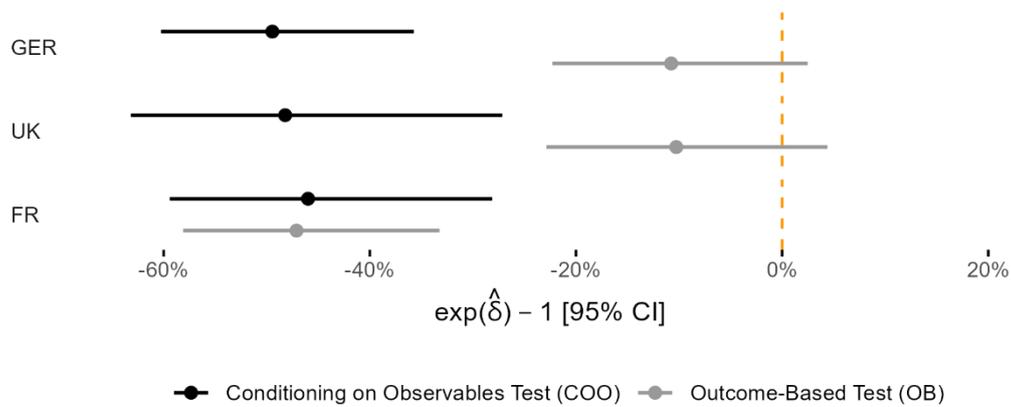

**Web Appendix B: Sensitivity Analysis for Impact of Measurement Error on Conclusion of Study A**

In Study A, we did not identify a statistically significant coefficient estimate of the *isAmazonBuyBox* indicator. This result could have occurred for several reasons: (i) because there is no true effect (the conclusion we draw in our study), (ii) because measurement error biases the coefficient estimates, or (iii) because our test lacks statistical power, thereby preventing us from rejecting the null even though there is a true effect ("type-2 error"). While we consider insufficient power an unlikely explanation because of the large number of observations and consistent results when pooling samples, we conduct three analyses to provide empirical evidence that the first explanation ("no true effect") is more likely than the others.

First, we discuss in which cases measurement error could occur and how often such cases are likely to arise. Second, we re-run our analysis, excluding observations where measurement error is likely to occur or controlling for it in the regression estimation. Finally, we show the robustness of the results by comparing those coefficient estimates to the original ones.

First, measurement error could be present in the time surrounding a buy box change since we estimate the impact of the previous day's last observation of the independent variable ($t-1$) on the focal day's dependent variable visibility ($t$). We measure visibility at a random time of the day $t$. Hence, in the most extreme case, a day's only and thereby last buy box change could happen at 0:01 am on day $t-1$, and visibility could be recorded on the next day ($t$) at 11:59 pm.

Accordingly, the time between a buy box change and the recording of visibility lies in the interval of (0, 48) hours. Higher values are unlikely since we use the previous day's last observation for the independent variables. Intuitively, measurement error could occur either on the day of the buy box change ($t$) or the day before ($t-1$) if the buy box changes again before the visibility measurement. In our three best-selling samples, a buy box change occurs, on average, every 10.1 days. Hence, we can assume that, on average, during at least 8.1 days of the 10.1 days without a recorded buy box change, our recorded buy box seller was the true buy box seller.

A reason for frequent buy box changes could be algorithmic price changes. Chen, Mislove, and Wilson (2016, p. 7) find that 2.4% of sellers in their dataset use algorithmic pricing. Further, in Figure 24, they show that for 65% of non-algorithmic sellers, the price never changes. In comparison, more than 95% of non-algorithmic sellers and more than 65% of algorithmic sellers have ten or fewer price changes in their observation period of 42 days—i.e., an average interval between price changes of at least 4.2 days.

Hence, we can assume that most algorithmic and non-algorithmic sellers change their prices at a significantly lower frequency than our daily measurement frequency. Accordingly, we argue that even if a measurement error occurs for products with very volatile prices, the impact of this likely random measurement error is very small due to the low number of products with highly volatile prices and, hence, does not materially affect our conclusions.

Second, we empirically quantify the potential impact of the above-described measurement error on Study A's conclusion (reject the null hypothesis of $\delta_{isAmazonBuyBox} = 0$ vs. fail to reject at the significance level of 5%) in Figure 8.

This visualization compares the original coefficient estimates (black) to the results from

- the original equation with an additional indicator for a buy box change at $t$
- the original equation with an additional indicator for a buy box change at $t$ and $t-1$,
- samples excluding the day of the buy box change,
- samples excluding both the day of the buy box change and the day before.

**Figure 8:** Study A—Quantifying the Impact of Potential Measurement Error on Coefficient Estimates and Standard Errors by Sample.

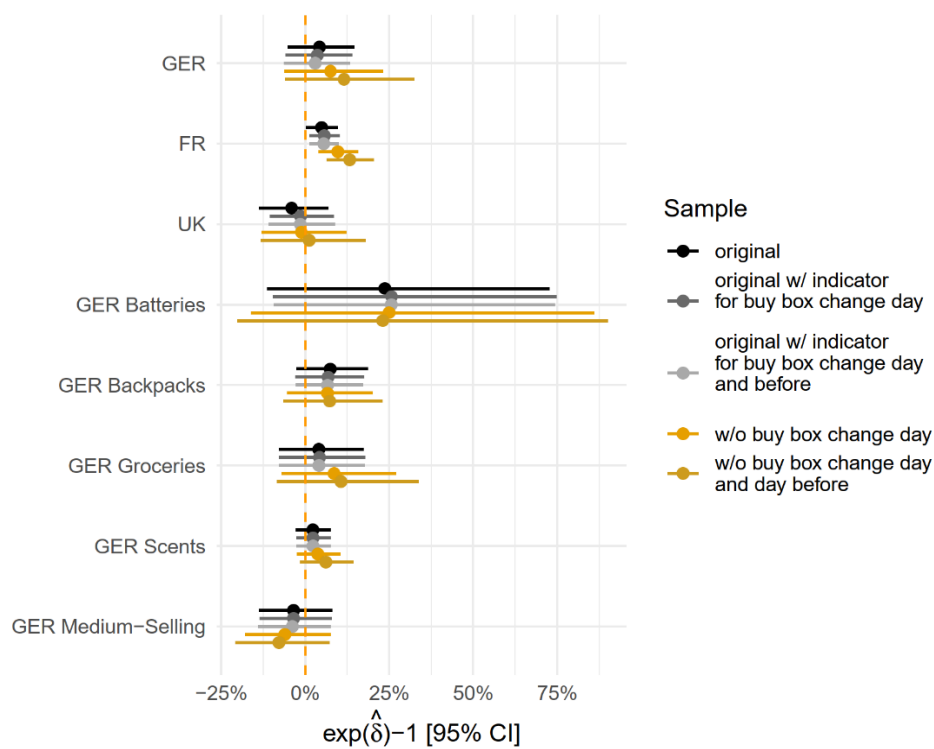

*Note: The figure compares the original coefficient estimates for each sample with coefficient estimates from four alternative specifications, removing days or controlling for days where measurement error is most likely.*

This comparison shows that the values of the coefficient estimates vary marginally, indicating that the original estimates could have been slightly biased. However, the confidence intervals overlap to a large degree, and every single conclusion on whether to reject the null hypothesis remains the same—apart from the French sample's coefficient, all estimates remain insignificantly different from zero at the 5% significance level. Hence, we conclude that measurement errors stemming from differences in measurement timing across the day are unlikely to change our conclusions.

# Web Appendix C: Simulation of Different Omitted Variable Types and Their Impact on Study A's Conclusion

The Conditioning-on-Observables (COO) test for self-preferencing can only yield an unbiased self-preferencing estimate if the regression modeling the recommendation as a function of protected and unprotected attributes does not suffer from omitted variable bias. More specifically, the coefficient of the protected attribute is only unbiased if we do not omit an unprotected attribute, which correlates both i) with the protected attribute and ii) with the recommendation.

Translated to the empirical setting of Study A, our regression in equation (7) needs to control for all unprotected attributes that are i) correlated with *isAmazon*, and ii) correlated with a product's organic visibility. Although we have no reason to believe that we omit any observable attribute satisfying those conditions, such unobservable attributes might still exist.

Web Appendix A already provides evidence for the robustness of our results by using the Outcome-based (OB) test as an alternative test for self-preferencing that does not require the same assumptions. This Web Appendix C presents another approach. It builds upon the idea that the inclusion of certain omitted variables would increase the self-preferencing estimate—the coefficient of our protected attribute, *isAmazon*—and others would decrease it. For certain settings, we can then rule out self-preferencing.

For example, suppose that we find a value of zero for the *isAmazon* coefficient, and all omitted variables, if included, would decrease the coefficient. In that case, self-preferencing cannot occur because the coefficient is zero, and the inclusion of omitted variables, if they even existed, would decrease, not increase the *isAmazon* coefficient. In other settings, such a conclusion is not possible. Suppose, for example, that we find a value of zero for the *isAmazon* coefficient, and all potentially omitted variables are likely to increase the coefficient's value. In that case, we cannot draw the same conclusion—the value of zero could either occur because (i) there are no omitted variables and, thus, no self-preferencing occurs or (ii) there are omitted variables that yield an underestimation of the self-preferencing coefficient so that its true value is larger than zero, indicating self-preferencing. After outlining the general idea, we now translate it into our specific setting.

## *"Amazon Advantage" Omitted Variable*

Suppose there is an omitted variable whose value is *higher* for Amazon-related products and that benefits a product's visibility. Higher means better so that the omitted variable represents an "Amazon advantage." In econometric terms, this omitted variable of the "Amazon advantage" type positively correlates with the protected attribute *isAmazon* and the dependent variable, organic *visibility*. On the Amazon marketplace, an omitted variable could be how well a seller keeps sufficient product quantities in stock. Amazon could "reward" sellers with high product availability by improving their products' visibility in the search results. However, Amazon has a reputation for a well-functioning supply chain and logistics services. Therefore, we suppose that Amazon is better at keeping stock of its products than third-party sellers, who might have more limited capabilities to keep products in stock. Hence, the potential omitted variable "in-stock percentage" would positively correlate with *isAmazon* and the dependent variable visibility.

So, how does omitting this variable affect the coefficient on *isPlatform*, our self-preferencing estimate? Assuming all else equal, the coefficient of the "Amazon advantage" variable would be positive due to its positive correlation with *visibility*. However, the "Amazon advantage" variable also positively correlates with *isAmazon*. Thus, the *isAmazon* and the "Amazon advantage" coefficients would *jointly* explain changes in the dependent variable visibility.

In this scenario, failing to control for the unobserved "Amazon advantage" variable would cause the coefficient of *isAmazon* to decrease. Hence, this scenario of omitted variable bias would lead to a negatively biased self-preferencing estimate.

Study A's estimates of *isAmazon* center around zero and fail to identify a statistically significant self-preferencing estimate in seven out of eight samples. This finding leads to the conclusion that we cannot provide much evidence *for* self-preferencing. Also, Study B estimates statistically significantly negative self-preferencing. Hence, even if our analysis suffered from an "Amazon advantage" omitted variable, we would still find no evidence of self-preferencing when correctly controlling for the "Amazon advantage" variable. Instead, failing to control for such a variable would let the self-preferencing estimate to become more strongly negative. In such a scenario, Amazon would treat its own products worse than its competitors' products. However, such "negative" self-preferencing would be of little regulatory concern, and third-party sellers would certainly not complain if their products received more visibility than they "deserved." Thus, an omitted "Amazon advantage" variable does not change the conclusion of our Study A that no self-preferencing occurred.

*"Amazon Disadvantage" Omitted Variable*

Conversely, assume that an omitted variable value is *lower* for Amazon-related products and benefits a product's visibility. In econometric terms, this omitted variable of the "Amazon disadvantage" type negatively correlates with the protected attribute *isAmazon* and positively with the dependent variable, organic *visibility*. Loosely speaking, such an attribute is an attribute that has higher values for third-party sellers (compared to Amazon) and benefits a product's visibility. Given Amazon's reputation for high-quality customer service and advanced logistics capabilities, it seems unlikely (although not impossible) that an omitted variable exists that satisfies those criteria.

Only in such a scenario, failing to control for an "Amazon disadvantage" variable would bias our self-preferencing estimate in a direction that could change our conclusion from finding "no evidence for self-preferencing" to "finding evidence for self-preferencing."

*Simulation Study Based on Study A*

In the following, we illustrate the effects of omitting either an "Amazon advantage" or an "Amazon disadvantage" variable by simulating such variables and then showing the impact on the value of *isAmazon*—our self-preferencing estimate.

For illustrative purposes, we consider Study A's pooled best-selling sample of products from the German, French, and British marketplaces, i.e., Model 0 of Table 5. We generate two new variables that serve as our "omitted" variables. The "Amazon Advantage" variable correlates highly positively with both the dependent variable *visibility* and *isAmazon* by constructing it as follows:

$$\text{AmazonAdvantage}_{it} = 15 \times \text{isAmazon}_{it} + \widetilde{VI}_{it} + v_{it} . \tag{10}$$

In this equation, *isAmazon*$_{it}$ refers to whether Amazon occupies the buy box for product $i$ on day $t$, $\widetilde{VI}_{it}$ is the product's organic visibility, and $v_{it}$ is a random error term, normally distributed with a zero mean and standard deviation of one. We multiply the *isAmazon* variable by 15 to produce meaningful level differences in the "Amazon advantage" variable.

Conversely, we construct the "Amazon disadvantage" variable by multiplying the *isAmazon* variable by negative 15 instead of positive 15:

$$\text{AmazonDisadvantage}_{it} = -15 \times \text{isAmazon}_{it} + \widetilde{VI}_{it} + v_{it} . \tag{11}$$

This construction yields the desired correlations, which we illustrate in Table 8. Both simulated variables correlate strongly with the dependent variable, *visibility* (= *organicVisibility*). While the "Amazon Advantage" variable correlates positively with isAmazon with 0.626, the "Amazon disadvantage" variable exhibits a negative correlation at - 0.578.

**Table 8:** Correlation Matrix of Simulated Amazon Advantage and Amazon Disadvantage Variables with Visibility and the Amazon Indicator

| *Variable* | organicVisibility | isAmazon | AmazonAdvantage | AmazonDisadvantage |
|---|---|---|---|---|
| organicVisibility | — | .048 | .804 | .782 |
| isAmazon |  | — | .626 | - .578 |
| AmazonAdvantage |  |  | — | .266 |
| AmazonDisadvantage |  |  |  | — |

Now, we suppose that our main analysis, displayed in Model 0 of Table 5, would suffer from an omitted variable, i.e., either the "Amazon advantage" or "Amazon disadvantage" variable. Table 9 shows how controlling for those variables affects the self-preferencing estimate—the coefficient of *isAmazonBuyBox (= isAmazon)*.

**Table 9:** Study A—Generalized Linear Model with Log-Link: Illustrating the Impact of Including a Simulated Omitted Variable "Amazon Advantage" or "Amazon Disadvantage" on the isAmazonBuyBox Coefficient.

|  | Dependent variable: organicVisibility | | |
|---|---|---|---|
| Country: | GER, FR, UK | GER, FR, UK | GER, FR, UK |
| Sample type: | best-selling | best-selling | best-selling |
| Model | (0) | (1) | (2) |
| *Variables* | | | |
| isAmazonBuyBox | .016 | -.132** | .276*** |
|  | (.029) | (.051) | (.053) |
| AmazonAdvantage |  | .014*** |  |
|  |  | (.002) |  |
| AmazonDisadvantage |  |  | .014*** |
|  |  |  | (.002) |
| *Fixed effects* | | | |
| Product | Yes | Yes | Yes |
| Date | Yes | Yes | Yes |
| No. of products | 1,344 | 1,344 | 1,344 |
| Observations | 270,482 | 270,482 | 270,482 |
| Pseudo R² | .713 | .736 | .736 |

*Notes: All models include the same control variables as in Table 5, but we omit their estimates for brevity. Model (0) reproduces the results from Model (0) in Table 5. Two-way (product and date) standard errors in parentheses. Significance Codes: \*\*\*: .01, \*\*: .05, \*: .1*

The results in Table 9 show that omitting an "Amazon advantage" variable, even though it leads to a biased self-preferencing estimate, does not threaten the conclusion that we do not find evidence for self-preferencing. Instead, the original non-significant self-preferencing estimate of the variable *isAmazonBuyBox* in Model (0) (corresponding to Model (0) in Table 5) becomes more negative when controlling for the simulated "Amazon advantage" variable. Hence, even if the original analysis in Model (0) suffered from an omitted "Amazon advantage" variable, our conclusion would remain the same, namely that we do not find evidence for self-preferencing.

Conversely, the other type of omitted variable—the "Amazon disadvantage" variable—would cause us to find evidence for self-preferencing erroneously. This mechanism becomes apparent when comparing Model (0) (identical to Model (0) in Table 5) to Model (2). There, when including the "Amazon disadvantage" variable, the coefficient increases and is now statistically significantly positive. Hence, if an "Amazon disadvantage" variable existed, representing an unprotected attribute that influences *visibility*, then we could erroneously conclude that there is evidence for self-preferencing, even though there was not.

In sum, this empirical illustration shows that only a certain type of omitted variable could bias our self-preferencing estimates. Specifically, only omitting an "Amazon disadvantage" variable, i.e., a variable that correlates positively with visibility but negatively with isAmazon, could threaten our conclusion of not finding evidence for self-preferencing. In other words, our self-preferencing estimates could only change to positive if an omitted variable that positively influences a product's visibility existed that is consistently higher for third-party sellers than for Amazon.

**Web Appendix D: Sensitivity Analysis for Impact of Dissimilar Products in Comparison Groups on Conclusion of Study B**

Study B estimates self-preferencing of Amazon Basics products compared to competing third-party private-label products. Therefore, we built 752 comparison groups of one Amazon Basics product and up to five comparable products across the three marketplaces in Germany, the United Kingdom, and France. Although the comparable products had to fulfill the same selection criteria as the respective Amazon Basics product, we cannot fully rule out that some of the selected products are not comparable to the Amazon Basics product. For example, our proxies for brand equity, seller and product ratings, might not be perfect. Therefore, we use additional criteria to select comparable products and examine whether the stricter selection impacts our results.

More precisely, we compute the ratios of the organic visibility of the Amazon Basics product to their comparison group's (third-party) product visibility and its inverse value. High values and values close to zero, respectively, indicate that the visibility of the Amazon Basics product differs strongly from its assigned competitors. So, we exclude all comparable products that exhibit a ratio above a cutoff $x$, or below $1/x$, respectively. We vary the (integer) cutoff from 1 to 30. Figure 9 and Figure 10 report the results of this sensitivity analysis for $x \in [1, 30]$, and compare those to the original estimates from Study B. Figure 9 visualizes the share of dropped comparison groups by the cutoff ratio.

**Figure 9:** Study B—Share of Dropped Comparison Groups by Cutoff for Comparison Groups' Visibility Ratio of Amazon Basics vs. Non-Amazon Basics Products and Vice Versa.

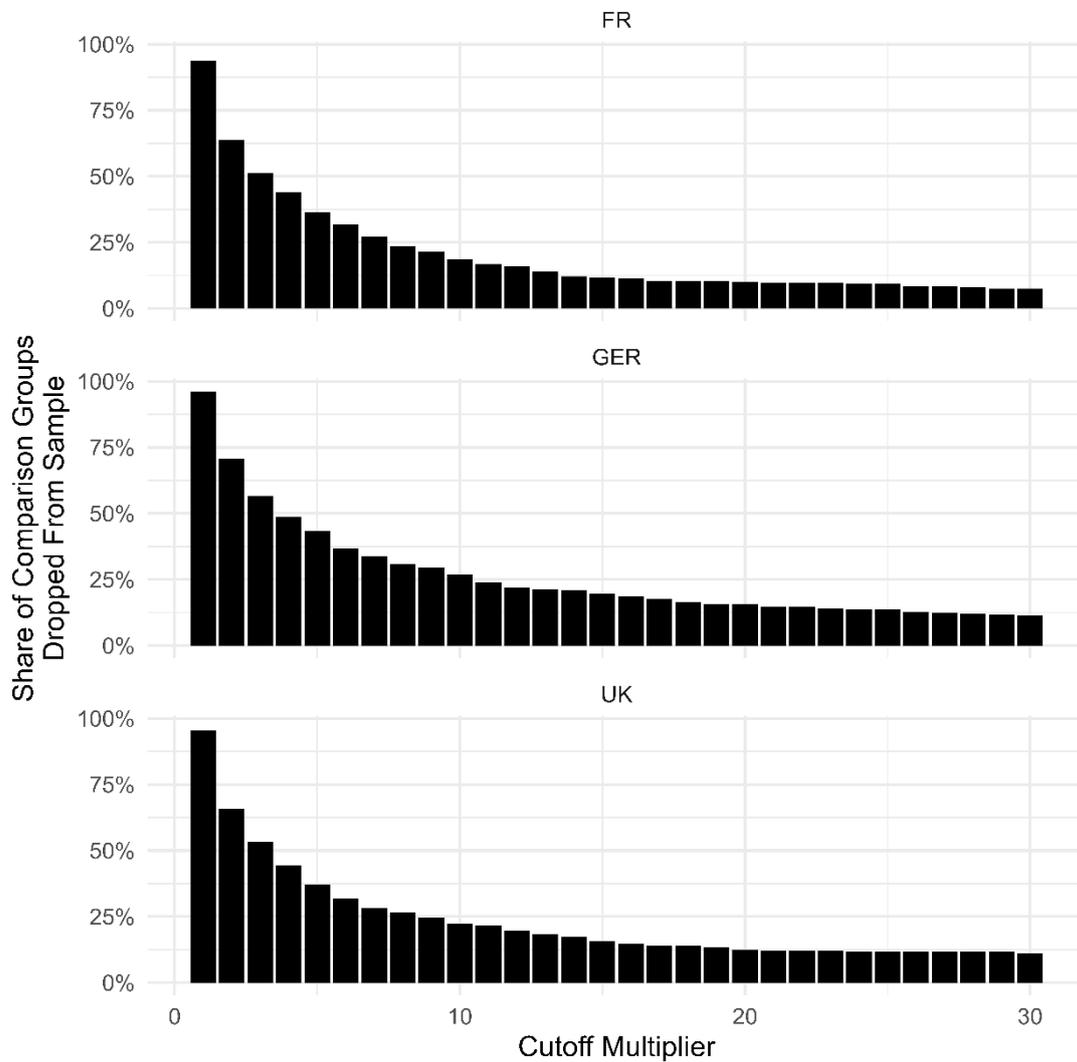

Figure 10 shows the results of the self-preferencing estimates for Amazon Basics products, depending on the different cutoff ratios. Overall, the results remain robust when excluding comparison groups with different visibility levels. The estimates in each of the three marketplaces remain substantially negative for all thresholds and remain of similar magnitude in the German marketplace. The estimates in the French and United Kingdom samples become slightly less negative than in the full sample but remain around -40%. For x > 30, the estimates remain almost unchanged. Hence, this sensitivity analysis gives us more confidence that the comparison groups are reasonably similar, and the large difference in self-preferencing estimates between the two studies is not an artifact of accidentally assigning products with vastly different visibility levels to individual comparison groups.

**Figure 10:** Study B—Sensitivity Analysis of Self-Preferencing Estimate $\hat{\delta}$ by Cutoff for Comparison Groups' Visibility Ratio of Amazon Basics vs. Non-Amazon Basics Products and Vice Versa.

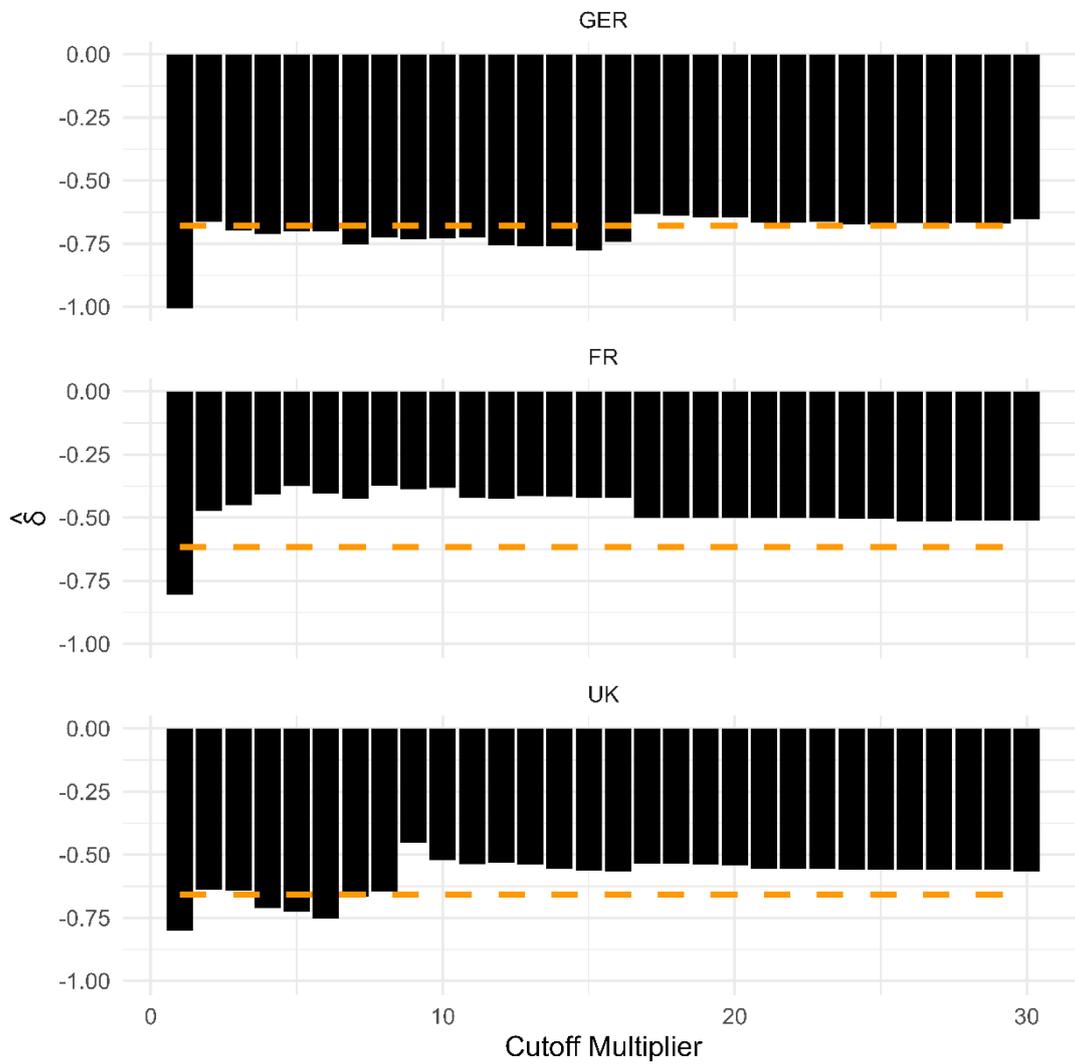

*Note: The orange dashed line visualizes the results without dropping any comparison groups, as shown in the main results of Study B in Table 7.*

## Web Appendix References

Chen, Le, Alan Mislove, and Christo Wilson (2016), "An Empirical Analysis of Algorithmic Pricing on Amazon Marketplace," in *Proceedings of the 25th International Conference on World Wide Web*, Montréal Québec Canada: International World Wide Web Conferences Steering Committee, 1339–49.

Reimers, Imke and Joel Waldfogel (2023), "A Framework for Detection, Measurement, and Welfare Analysis of Platform Bias," NBER Working Paper 31766.